\documentclass[a4paper,11pt]{article}
\pdfoutput=1 

\usepackage{jcappub} 
\usepackage[toc,page]{appendix}
\usepackage{subcaption}
\usepackage{amsmath}
\usepackage{graphicx}                
\usepackage{multirow}
\usepackage[T1]{fontenc} 
\usepackage[table]{xcolor}
\usepackage{float}

\usepackage{orcidlink}

\definecolor{verde}{rgb}{0,0.5,0}

\hypersetup{pdftitle={Closing in on alpha-attractors}
}
\usepackage{bm}

\title{\boldmath 
Closing in on $\alpha$-attractors}

\author[a,b]{Laura Iacconi,\,$^{\text{\orcidlink{0000-0002-1152-3056}}}$}
\author[c]{Sukannya Bhattacharya,\,$^{\text{\orcidlink{0000-0003-4464-5874}}}$}
\author[c]{Matteo Fasiello,\,$^{\text{\orcidlink{0000-0002-2532-5202}}}$}
\author[b]{David Wands\,$^{\text{\orcidlink{0000-0001-9509-8386}}}$}
\affiliation{$^{a}$Astronomy Unit, Queen Mary University of London, \\ Mile End Road, London, E1 4NS, U.K.}
\affiliation{$^{b}$ Institute of Cosmology \& Gravitation, University of Portsmouth,\\ Burnaby Road, Portsmouth, PO1 3FX, U.K.}
\affiliation{$^{c}$ Instituto de F\'{i}sica T\'{e}orica UAM-CSIC, c/ Nicol\'{a}s Cabrera 13-15, 28049, Madrid, Spain}

\emailAdd{l.iacconi@qmul.ac.uk}
\emailAdd{sukannya.bhattacharya@ift.csic.es}
\emailAdd{matteo.fasiello@csic.es}
\emailAdd{david.wands@port.ac.uk}

\abstract{
Recent observations of cosmic microwave background (CMB) anisotropies combined with large-scale structure may point towards higher values of the scalar spectral index, $n_s$. This puts previously preferred inflationary models, such as $\alpha$-attractors, in tension with the new measurements.
Pending a resolution of the tension between BAO parameters as determined by CMB datasets and those determined by DESI, we explore in this work the large-$n_s$ regime of $\alpha$-attractor T-models. 
We show that some T-models can self-consistently produce an extended reheating stage with a stiff equation of state $(\bar w>1/3)$, which allows values for $n_s$ closer to unity.
We employ constraints from P-ACT-LB-BK18 data to illustrate what large-$n_s$ observations might imply for T-models with monomial potentials. 
We show that the $n_s$ measurement yields an upper limit on $\alpha$ that is stronger than the one from the tensor-to-scalar ratio only.
We find that $n_s$ is maximised for $\alpha\sim1$, therefore the seven Poincar\'e models are well placed to deliver large $n_s$.
However, the ability of a stiff reheating stage to increase the compatibility of T-models with large-$n_s$ measurements saturates as $\bar{w}\to1$.
Thanks to this effect, we establish that the largest $n_s$ that monomial T-models can produce is $n_s=0.9682$. 
T-models are therefore highly predictive in the large-$n_s$ regime and our result provides, under the assumption of perturbative reheating, a benchmark which could be used in the future to rule out monomial T-models.
}

\begin{document}
	\maketitle
	\flushbottom
	
\section{Introduction}
\label{sec: introduction}

Single-field, slow-roll inflation provides a simple scenario in which to describe the origin of the large-scale structure of our Universe from microscopic, quantum fluctuations at very early times. An almost-de Sitter expansion during inflation naturally gives rise to an almost-scale invariant distribution of primordial density perturbations, with the deviation from scale-invariance controlled by dimensionless slow-roll parameters, $\epsilon_i\sim0.01$~\cite{Liddle:2000cg,Ellis:2023wic,ParticleDataGroup:2024cfk}.

Data from current cosmic microwave background (CMB) experiments
\textit{Planck}~\cite{Planck:2018jri}, ACT~\cite{ACT:2025fju} and SPT-3G~\cite{SPT-3G:2025bzu}, yield consistent measurements of the primordial scalar spectral tilt under the assumption of a $\Lambda \text{CDM}$ Universe. 
These are 
\begin{align}
    \text{\textit{Planck} 2018}&: \quad n_s = 0.9649 \pm 0.0042 \quad  (68\%) \,\text{C.L.} \;,\\
    \text{ACT DR6}&: \quad n_s= 0.9666 \pm 0.0077 \quad  (68\%) \,\text{C.L.} \;,\\
    \text{SPT-3G D1}&: \quad n_s= 0.949 \pm 0.013 \quad  (68\%) \,\text{C.L.} \;.  
\end{align}
Once BAO DESI data are added, CMB datasets show a preference for larger $n_s$ values, with the largest being produced by ACT DR6 and SPT-3G.
This is due to the correlation between $n_s$ and BAO parameters, and the tension between BAO parameters as determined from CMB and DESI data~\cite{SPT-3G:2025bzu,Ferreira:2025lrd}. 
To illustrate this effect, we focus on ACT DR6 data, and quote the results obtained for the $\Lambda\text{CDM}+r$ model. 
We employ $\Lambda\text{CDM}+r$ because cosmological inflation predicts the existence of gravitational waves (GWs) of primordial origin, whose amplitude is parametrised by the tensor-to-scalar ratio, $r$.
By using ACT DR6, \textit{Planck} data with $\ell < 1000$ in the TT channel and $\ell < 600$ in TE/EE, the \textit{Planck} measurement of the optical depth to reionization, CMB lensing data, DESI Year-1 BAO data and BICEP-Keck 2018 data one obtains~\cite{ACT:2025tim}
\begin{align}
\label{eq:ns P-ACT-LB-BK18}
    n_s = 0.9741 \pm 0.0033 \quad &(68\% \text{C.L.}) \;, \\
\label{eq:r P-ACT-LB-BK18} 
    r < 0.038 \quad &(95\% \text{C.L.}) \;. 
\end{align}
As in Ref.~\cite{ACT:2025tim}, we label this data combination as P-ACT-LB-BK18. 

On one hand, by using the new measurement~\eqref{eq:ns P-ACT-LB-BK18} to constrain inflation one might conclude that once-preferred-by-the-data single-field slow-roll models (with standard reheating), such as the Starobinsky model~\cite{Starobinsky:1980te}, are now disfavoured at more than $2\sigma$. 
On the other hand, before applying Eq.~\eqref{eq:ns P-ACT-LB-BK18} to constrain inflationary models, the tension between CMB data and DESI BAO data must be resolved~\cite{Ferreira:2025lrd}. 
Nevertheless, in case this preference for large $n_s$ will be confirmed, e.g., by future CMB data, it is important to explore the large-$n_s$ regime of inflationary models. 

In this work, we focus on single-field monomial T-models, defined by the inflaton potential~\cite{Kallosh:2013hoa}
\begin{equation}
\label{eq: T-model potential}
    V(\phi)=V_0\tanh^p{\left( \frac{\phi}{\sqrt{6\alpha} M_\text{Pl}}\right)} \;,
\end{equation} 
where $p$ is an even integer satisfying $p\geq 2$. 
T-models belong to a class of cosmological $\alpha$-attractors displaying an exponential approach to the large-field potential plateau. 
Generally, $\alpha$--attractors~\cite{Kallosh:2013hoa, Kallosh:2013daa, Ferrara:2013rsa, Kallosh:2013pby, Kallosh:2013lkr,Kallosh:2013maa, Kallosh:2013tua, Kallosh:2013yoa} provide a compelling mechanism for acceleration in the early universe for several reasons. 
They can be embedded in supergravity theories and M-theory (see e.g. Ref.~\cite{Ferrara:2013rsa}), and their predictions lie close to the center of \textit{Planck} observational bounds on the primordial power spectra~\cite{Planck:2018jri}.
Furthermore, a large fraction of the parameter space of $\alpha$-attractors  leads to universal predictions for large-scale observables that are broadly independent of the details of the inflationary potential~\cite{Kallosh:2013hoa}.

T-models yield predictions for $n_s$ closer to unity when reheating has a stiff equation of state, $\bar w>1/3$, and lasts sufficiently long.
This is due to the fact that in this case, reheating increases the number of e-folds elapsed from the horizon crossing of the CMB pivot scale to the end of inflation, see Eqs.~\eqref{eq: DN CMB} and~\eqref{eq: ns universal}. 
Importantly, $\bar w$ is not a free parameter, but rather it is determined by the potential parameter $p$, with $p>4$ yielding $\bar w>1/3$. 
In this work we therefore explore the large-$p$ limit of T-models. 

The ability of reheating to bring $\alpha$-attractor predictions closer to ACT DR6 and SPT-3G measurements of $n_s$ was pointed out in Refs.~\cite{Drees:2025ngb, Haque:2025uri, Haque:2025uga, German:2025ide, Ellis:2025zrf}. 
In Sec.~\ref{sec: discussion} we will detail how our work compares with and goes beyond previous studies.

\medskip
\textit{Content:} 
this paper is organised as follows. 
In Sec.~\ref{sec: inflation and reheating} we describe how CMB observables are connected to the dynamics of the inflaton (Sec.~\ref{sec: constraining inflation with CMB observations}) and discuss the contribution of reheating in detail (Sec.~\ref{sec: reheating intro}). 
In Sec.~\ref{sec: effective equation of state parameter} we show how the equation-of-state parameter is determined by the  parameter $p$ in the potential.  
In Sec.~\ref{sec: duration of reheating} we present the upper bounds on the duration of reheating. 
We require that reheating is complete either (i) at temperatures above $1\,\text{TeV}$ in order to allow for baryogenesis to take place, or (ii) just before Big Bang nucleosynthesis (BBN).
For models with stiff reheating, we also implement a second constraint on the duration of reheating, stemming from the BBN upper limit on primordial GWs. 
In Sec.~\ref{sec: inflationary predictions} we present $n_s$ and $r$ predictions for T-models with even $p\in[2,\,20]$ and $10^{-4}\leq \alpha\leq 20$. 
We describe how these are obtained in Sec.~\ref{sec: how ns and r are computed}, and compare them with the posterior from P-ACT-LB-BK18 in the $(n_s,\,r)$ plane in Sec.~\ref{sec: predictions vs PACT}. 
In Sec.~\ref{sec: Parameter space constraints from P-ACT-LB-BK18} we explore what CMB data can tell us about T-models if -- pending tensions being resolved-- future CMB measurements will support the preference
brought on by new CMB datasets for $n_s$ larger than that determined by \textit{Planck}. 
To this end, we employ the $(n_s,\,r)$ P-ACT-LB-BK18 posterior to constrain the $(\alpha,\,p)$ parameter space. 
In Sec.~\ref{sec: how far can we push ns in these models} we determine what $n_s$ measurement might allow us in future to rule out T-models by computing the largest $n_s$ that T-models can produce.
We close in Sec.~\ref{sec: discussion} with a detailed discussion of our results and a comparison with the existing literature. 

\medskip
\textit{Conventions:} Throughout this work, we consider a spatially-flat Friedmann--Lema\^{i}tre--Robertson--Walker universe, with line element $\text{d}s^2=-\text{d}t^2+a^2(t)\delta_{ij}\text{d}x^i\text{d}x^j$, where $t$ denotes cosmic time and $a(t)$ is the scale factor. The Hubble rate is defined as $H\equiv {\dot a}/{a}$,  where a derivative with respect to cosmic time is denoted by $\dot f \equiv {\mathrm{d}f}/{\mathrm{d}t}$. The number of e-folds of expansion is defined as $N\equiv \int\,H(t)\mathrm{d}t$ and $f'\equiv {\mathrm{d}f}/{\mathrm{d}N}$. 
The reduced Planck mass is $M_\text{Pl}\equiv(8\pi G_N)^{-1/2}$.

\section{Cosmological inflation and reheating}
\label{sec: inflation and reheating}

Given a single-field potential, $V(\phi)$, such as the T-model one~\eqref{eq: T-model potential}, the equations of motion for the background evolution are 
\begin{align}
\label{eq: single field inflaton eom}
    \ddot \phi+3H\dot \phi+V_\phi&=0\;, \\
\label{eq: phi dot with H dot}
    \dot H+ \frac{1}{2} \frac{\dot \phi^2}{M_\text{Pl}^2} &=0 \;,
\end{align}
where $V_\phi\equiv{\mathrm{d}V(\phi)/\mathrm{d}\phi}$, and the system obeys the Friedmann equation
\begin{equation}
\label{eq: Friedmann single field}
    H^2=\frac{1}{3 M_\text{Pl}^2}\left[\frac{1}{2}\dot \phi^2 +V(\phi)\right] \;.
\end{equation}
To obtain the results presented in the upcoming sections, we numerically solve Eqs.~\eqref{eq: single field inflaton eom} and~\eqref{eq: phi dot with H dot} by using the T-models potential~\eqref{eq: T-model potential} for a range of values of the potential parameters $p$ and $\alpha$. 
Note that, in practice, instead of using cosmic time, $t$, we employ the number of e-folds of expansion, defined by $\mathrm{d}N\equiv H(t)\mathrm{d}t\, $. 

Assuming that the inflaton starts evolving from large and positive field value, $\phi_\text{in}\gg \sqrt{6\alpha} \,M_\text{Pl}$, the inflaton dynamics is well described within the \emph{slow-roll approximation}. 
This assumes that the inflaton is overdamped ($|\ddot \phi| \ll 3H|\dot \phi|$) and slowly rolling ($\dot \phi^2\ll V(\phi)$) down the slope of its potential, which for  $\phi\gg \sqrt{6\alpha} \,M_\text{Pl}$ displays a plateau.
In this regime an inflationary stage can indeed take place, with $H^2\approx V(\phi)/(3M_\text{Pl}^2)$. 
One can characterise slow-roll inflation by introducing a hierarchy of \emph{slow-roll parameters}, defined as
\begin{equation}
\label{eq: slow roll parameters definition}
    \epsilon_0 \equiv M_\text{Pl} H^{-1},  \; \epsilon_{i+1} \equiv \frac{\dot \epsilon_i}{H \epsilon_i} \; \forall \,i\geq 0 \;. 
\end{equation}
Accelerated expansion requires $\epsilon_1<1$; slow-roll inflation corresponds to $|\epsilon_i|\ll 1 \; \forall \,i\geq 1$.
It is during slow-roll inflation that large-scale modes of vacuum fluctuations are produced and stretched beyond the cosmological horizon.  
Upon horizon re-entry, these provide the seeds for CMB temperature fluctuations. 
We discuss more in details how the inflaton dynamics gets imprinted in CMB observables in Sec.~\ref{sec: constraining inflation with CMB observations}. 

When the inflaton leaves the plateau, it gains momentum, exits the slow-roll regime and starts oscillating around the minimum of its potential. 
Inflation ends when the slow-roll parameters become of order unity. 
When numerically solving Eqs.~\eqref{eq: single field inflaton eom} and~\eqref{eq: phi dot with H dot}, we define the end of inflation as the time when $\epsilon_1$ becomes larger than unity.

Once the inflaton oscillates about the minimum of its potential, the inflaton, and/or its decay products, must decay into Standard Model particles, which rapidly thermalise. 
The process describing the energy transfer from the inflaton sector to ordinary matter goes by the name of \emph{reheating}~\cite{Kofman:1997yn}, which we discuss further in Sec.~\ref{sec: reheating intro}. 
There, we will relate the potential parameter $p$, see Eq.~\eqref{eq: T-model potential}, to the effective equation-of-state parameter during reheating. 

\subsection{Constraining inflation with CMB observations}
\label{sec: constraining inflation with CMB observations}

Cosmological inflation not only realises the accelerated background expansion needed to solve the main issues of the standard hot Big Bang model, but also explains the origin of scalar perturbations which later evolve into the large-scale structure we observe today. 
For concreteness, in the spatially-flat gauge these correspond to vacuum fluctuations of the inflaton field, $\phi(\mathbf{x}, \, t) \equiv \bar \phi(t) + \delta \phi(\mathbf{x}, \, t)$, where $\bar \phi(t)$ is the background field value obtained by solving Eqs.~\eqref{eq: single field inflaton eom} and~\eqref{eq: phi dot with H dot}.  
Each Fourier mode of $\delta \phi(\mathbf{x}, \, t)$, characterised by comoving wavenumbver $k$, oscillates while sub-horizon $(k\gg aH)$, and freezes into a classical perturbation after horizon crossing. 
The time of horizon crossing, $N_k$, is defined by $k\equiv (aH)|_{N=N_k}$. 
By counting the number of e-folds separating $N_k$ from the end of inflation, this can be translated into $\Delta N_k\equiv N_\text{end} - N_k$. 

Observations of the CMB are able to constrain the statistics of large-scale primordial perturbations, produced approximately $50-60$ e-folds before the end of inflation. 
More precisely, the number of e-folds separating the horizon crossing of the CMB pivot scale $k_\text{CMB}=0.05\,\text{Mpc}^{-1}$ from the end of inflation is
    \footnote{In Eq.~\eqref{eq: DN CMB} we have fixed the number of relativistic degrees of freedom at the end of reheating, $g_{\star,T_\text{rh}}=106.75$, where $T_\text{rh}$ is the temperature at the end of reheating. 
    In Secs.~\ref{sec: inflationary predictions} and~\ref{sec: how far can we push ns in these models} we let reheating last up to just before Big Bang nucleosynthesis takes place, in which case we use $g_{\star,T_\text{rh}}=10.75$ and the numerical factor in Eq.~\eqref{eq: DN CMB} is 61.1 instead.}
\begin{equation}
\label{eq: DN CMB}
\begin{split}
    \Delta N_\text{CMB} &\equiv N_\text{end} - N_\text{CMB} \\
    & \simeq 60.9 + \frac{1}{4} \ln{\left(\frac{V_\text{CMB}^2}{\rho_\text{end} M_\text{Pl}^4} \right)}  -\frac{1-3\bar w}{4} \Delta N_\text{rh} \;. 
\end{split}
\end{equation}
Here, $V_\text{CMB}$ is the value of the inflaton potential when the CMB mode crossed the horizon, $\rho_\text{end}$ is the energy density at the end of inflation, $\bar w$ is the effective equation-of-state parameter of reheating, and $\Delta N_\text{rh}$ is the duration of reheating measured in e-folds.  

Primordial scalar perturbations produced within canonical, single-field inflation are Gaussian to a very good level~\cite{Maldacena:2002vr}. 
They can therefore be characterised in terms of their two-point correlation function, whose Fourier transform is the scalar power spectrum, $\mathcal{P}_\zeta(k)$. 
At leading order in the slow-roll approximation, $\mathcal{P}_\zeta(k)$ is given in terms of the Hubble rate, $H$, and $\epsilon_1$ as \cite{ParticleDataGroup:2022pth}
\begin{equation}
\label{eq: slow roll power spectrum}
    \mathcal{P}_\zeta(k)=\frac{H^2}{8\pi^2 M_\text{Pl}^2 \epsilon_1} \Big|_{k=aH} \;,
\end{equation}
where this expression is evaluated for each $k$ mode soon after horizon crossing. 
Due to inflation's deviation from perfect de Sitter expansion, the primordial spectrum $\mathcal{P}_\zeta(k)$ is not exactly scale-invariant.  

For the purpose of comparing inflationary predictions to current CMB data, it is useful to parametrise the scalar power spectrum~\eqref{eq: slow roll power spectrum} on large scales with a simple power-law expression \cite{Planck:2018jri}, 
\begin{equation}
\label{eq: P zeta power law}
    \mathcal{P}_\zeta(k)=\mathcal{A}_s \left(\frac{k}{k_\text{CMB}} \right)^{n_s-1} \;,
\end{equation}
where $\mathcal{A}_s$ and $n_s-1$ are the amplitude and spectral tilt of $\mathcal{P}_\zeta(k)$ respectively, defined at $k_\text{CMB}$.
CMB experiments are compatible with primordial fluctuations with amplitude  $\approx 10^{-5}$, and slightly red-tilted power spectrum, see Eq.~\eqref{eq:ns P-ACT-LB-BK18}.

During cosmological inflation, tensor perturbations are also produced.
At leading order in slow-roll, the universally present contribution to the power spectrum of primordial tensor modes is 
\begin{equation}
    \label{eq: power spectrum of tensor modes}
    \mathcal{P}_t(k) = \frac{2H^2}{\pi^2 M_\text{Pl}^2} \Big|_{k=aH} \;.  
\end{equation}
The corresponding amplitude of primordial tensor models at $k_\text{CMB}$, $\mathcal{A}_t = {2H^2}/{(\pi^2M_\text{Pl}^2)}|_{k=k_\text{CMB}}$, is usually constrained relative to the scalar one by introducing the tensor-to-scalar ratio
\begin{equation}
    \label{eq: r definition}
    r \equiv \frac{\mathcal{A}_t}{\mathcal{A}_s} \;. 
\end{equation}
Gravitational waves of primordial origin have not yet been detected in B-mode polarisation of the CMB,  so that $r$ is constrained from above, see Eq.~\eqref{eq:r P-ACT-LB-BK18}. 

For a given inflationary model, one can compute the predictions for $\mathcal{A}_s$, $n_s$ and $r$ by evaluating the corresponding slow-roll expressions $\Delta N_\text{CMB}$ e-folds before the end of inflation.
(See Refs.~\cite{Auclair:2022yxs, Ballardini:2024irx} for expressions up to third-order in the slow-roll expansion.)   
We will present results for $n_s$ and $r$ computed for the T-models under analysis in Sec.~\ref{sec: inflationary predictions}. 
First though, we must address the reheating stage, as $\Delta N_\text{CMB}$ crucially depends on its duration and equation-of-state parameter, see Eq.~\eqref{eq: DN CMB}.

\subsection{Reheating}
\label{sec: reheating intro}
During reheating, the energy density decreases from its value at the end of inflation, $\rho_\text{end}$, to $\rho_\text{rh}$, when the inflaton, and/or its decay products, have decayed into Standard Model particles and these are thermalised. 
Note that while non-perturbative mechanisms may play an important role during reheating~\cite{Allahverdi:2010xz,Lozanov:2019jxc}, we will consider perturbative reheating, as modelling non-perturbative dynamics is beyond the scope of this work. 

The reheating stage can be characterised in terms of three quantities, two of which are independent. 
First, the energy density of radiation when reheating is complete, 
\begin{equation}
\label{eq: rho rh with T rh}
    \rho_\text{rh} \equiv \rho(T_\text{rh}) = \frac{\pi^2}{30} \, g_{\star,T_\text{rh}} \, {T_\text{rh}}^4 \;,
\end{equation}
where $T_\text{rh}$ is the temperature of radiation at the end of reheating and $g_{\star,T}$ is the number of effective relativistic degrees of freedom at temperature $T$.
Second, the effective equation-of-state parameter during reheating, $\bar w$. 
Third, the duration of reheating computed in terms of e-foldings of expansion after the end of inflation, $\Delta N_\text{rh}\equiv N_\text{rh}-N_\text{end}$ (here $N_\text{rh}$ marks the end of reheating). 
This can be expressed in terms of $\rho_\text{rh}$ and $\bar w$ as
\begin{equation}
    \label{eq: Delta N rh def}
    \Delta N_\text{rh} \equiv \frac{1}{3(1 + \bar w)}\log{\left( \frac{\rho_\text{end}}{\rho_\text{rh}}\right)} \;.
\end{equation}
Thanks to Eq.~\eqref{eq: Delta N rh def}, we fully characterise reheating in terms of the two dimensionless quantities, $\bar w$ and $\Delta N_\text{rh}$.
As we shall see, $\bar w$ is a theory-determined parameter, being related to the potential shape parameter, $p$. 
The duration $\Delta N_\text{rh}$ is bounded from above, but it is otherwise undetermined, and should therefore be treated alongside the theoretical parameters $p$ and $\alpha$. 

Taking the reheating phase into account is essential in order to correctly connect the dynamics of inflation with observations~\cite{Martin:2014nya}. 
Eq.~\eqref{eq: DN CMB} shows that its duration and its effective equation of state enter the computation of $\Delta N_\text{CMB}$, and have therefore an impact on CMB observables.
See Ref.~\cite{Mishra:2021wkm} for a discussion of how reheating can break degeneracies in the $n_s$ and $r$ predictions.

\subsubsection{Effective equation-of-state parameter}
\label{sec: effective equation of state parameter}

The effective equation-of-state parameter $\bar w$ is determined by the shape of the inflationary potential about its minimum, which for the T-models~\eqref{eq: T-model potential} we consider is 
\begin{equation}
\label{potential small phi}
    V(\phi)\sim \left(\frac{\phi}{\sqrt{6\alpha} \, M_\text{Pl}}\right)^p\;\;\; \text{for} \;|\phi|\ll \sqrt{6\alpha} \, M_\text{Pl}  \;. 
\end{equation}
This yields~\cite{Turner:1983he}
\begin{equation}
\label{eq: equation of state parameter as function of p}
    \bar w \approx \frac{p-2}{p+2} \;.
\end{equation}
In other words, the choice of inflationary potential determines the reheating equation-of-state parameter, and the two should not be treated as independent. 

From Eq.~\eqref{eq: equation of state parameter as function of p} one sees that when the inflaton can be described as an oscillating massive field ($p=2$), the background density redshifts like pressureless matter ($\bar w=0$), and when the potential is approximated with a quartic function ($p=4$), the background behaves as radiation ($\bar w=1/3$).
Instead, when $p \geq 6$ the reheating stage has a stiff equation of state $(\bar w>1/3)$, with asymptotic behavior $\bar w\to 1$ when $p\to \infty$. 

By considering the coefficient of the third term on the right-hand side of Eq.~\eqref{eq: DN CMB} one sees that depending on whether $\bar w$ is stiff or not, reheating contributes negatively $(p\geq 6 \to \bar w>1/3)$ or positively $(p<4\to \bar w<1/3)$ to $\Delta N_\text{CMB}$, thereby decreasing or increasing its value. 
When $p=4$ reheating has no impact on $\Delta N_\text{CMB}$ as the transition to radiation domination is effectively instantaneous. 
Stiff reheating will therefore bring the scalar spectral tilt closer to unity, see Eq.~\eqref{eq: ns analytic}.
In Sec.~\ref{sec: inflationary predictions} we will investigate the extent of this effect. 

Before moving on, let us comment on the time-dependence of the (instantaneous) equation-of-state parameter, $w=P/\rho$.  
At the end of inflation, $w$ transitions from $w \approx -1$ to $w>-1$, and $\rho$ decreases below $\rho_\text{end}$.
Of course, $w$ is not constant during reheating.
The inflaton oscillations are imprinted in its time-dependence, and overall $w$ grows from $-1$ at the end of inflation to $1/3$ at the onset of radiation domination. 
This complex dynamics is usually modeled in terms of an effective equation-of-state parameter, defined in Eq.~\eqref{eq: Delta N rh def}, which we will label here $w_\text{fit}$. 
In other words, for an explicit inflation model and a given value of $\Delta N_\text{rh}$, the two numerically-obtained points $\left(N_\text{end},\, \rho_\text{end}\right)$ and $\left(N_\text{rh},\,\rho_\text{rh}\right)$ can be fitted with the function in Eq.~\eqref{eq: Delta N rh def}, whose slope yields the parameter $w_\text{fit}(\Delta N_\text{rh})$. 
For $\rho_\text{th}\ll \rho_\text{end}$ (or $\Delta N_\text{rh}\gg 1$) the fitted value $w_\text{fit}(\Delta N_\text{rh})$ asymptotes to $\bar w$ in Eq.~\eqref{eq: equation of state parameter as function of p}.
We investigate the time-dependence of $w$ in Appendix~\ref{app: theoretical error from w time dependence} (see Fig.~\ref{fig:numerical w}), where 
we assess the magnitude of the theoretical error introduced in the computation of $n_s$ due to the use of Eq.~\eqref{eq: equation of state parameter as function of p} instead of the fitted value $w_\text{fit}(\Delta N_\text{rh})$. 
For two representative models with small and large $p$, we find that such error is always within $\sim1\%$ of the current observational uncertainty, see Eq.~\eqref{eq:ns P-ACT-LB-BK18}. 
This error should be considered alongside other sources of uncertainty in the computation of inflationary predictions, such as the error made in using Eq.~\eqref{eq: ns analytic} to compute $n_s$ instead of the second-order (or higher) slow-roll predictions (see Sec.~\ref{sec: how ns and r are computed}). 
Our analysis in Appendix~\ref{app: theoretical errors} shows that these two sources of theoretical uncertainty produce comparable errors, both well below the current experimental one, $\sigma(n_s)$. 
For this reason, we will employ Eq.~\eqref{eq: equation of state parameter as function of p} to determine the effective equation-of-state parameter.

\subsubsection{Duration of reheating} 
\label{sec: duration of reheating}

The information on reheating obtained from large-scale observations of the CMB is not enough to fully determine key information on reheating~\cite{Martin:2024qnn}, such as its duration ($\Delta N_\text{rh}$) and equation of state ($\bar w$).
Nevertheless, there are physically-motivated upper bounds on $\Delta N_\text{rh}$, on which we will rely to produce T-models inflationary predictions in Secs.~\ref{sec: inflationary predictions} and~\ref{sec: how far can we push ns in these models}.  

\subsubsection*{Temperature bound: $\bm{\Delta N_\text{rh, max 1}}$}
Reheating must be complete before the onset of Big Bang nucleosynthesis (BBN), $T_\text{BBN}=0.1\, \text{MeV}$. 
By assuming baryogenesis takes place after reheating is complete and considering that it requires physics beyond the standard model~\cite{Mukhanov:2005sc}, we place an upper bound on $\Delta N_\text{rh}$ by requiring $T_\text{rh}\geq 1 \, \text{TeV}$. 
The energy density at the end of reheating is therefore
\begin{equation}
\label{eq: first bound on duration of reheating}
    \rho_\text{rh} \geq \rho_\text{rh, min 1} \equiv \frac{\pi^2}{30}\; g_{\star,1\, \text{TeV}} \; \left( 1\,\text{TeV}\right)^4  \;. 
\end{equation}
Since in this scenario the QCD phase transition happens after reheating is complete $(T_\text{QCD}=150\,\text{MeV})$, we substitute $g_{\star,1\, \text{TeV}}=106.75$ in Eq.~\eqref{eq: first bound on duration of reheating}. 
If, at the completion of reheating, degrees of freedom beyond those of the standard model were present, $g_{\star,T_\text{rh}}$ could be much larger. 
Nevertheless, due to $\Delta N_\text{rh}$ being only logarithmically-sensitive to $\rho_\text{rh}$, our results are minimally affected by changes in $g_{\star,T_\text{rh}}$. 

For models with even values of $p\in [4,\,20]$ and $10^{-4}\leq \alpha\leq 20$, we numerically compute the energy density at the end of inflation, $\rho_\text{end}$, and by using Eqs.~\eqref{eq: first bound on duration of reheating},~\eqref{eq: Delta N rh def} and~\eqref{eq: equation of state parameter as function of p}, we find the associated maximum duration, $\Delta N_\text{rh, max 1}$.
\begin{figure}
\centering
\captionsetup[subfigure]{justification=centering}
    \begin{subfigure}{.48\textwidth}
        \includegraphics[width=\textwidth]{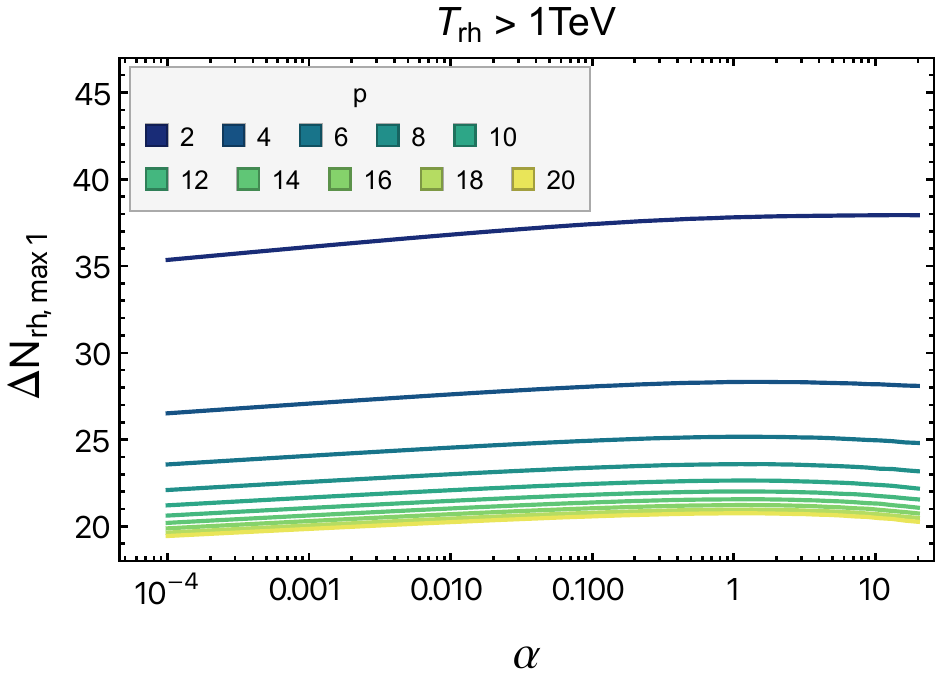}
    \end{subfigure}
    \begin{subfigure}{.48\textwidth}
        \includegraphics[width=\textwidth]{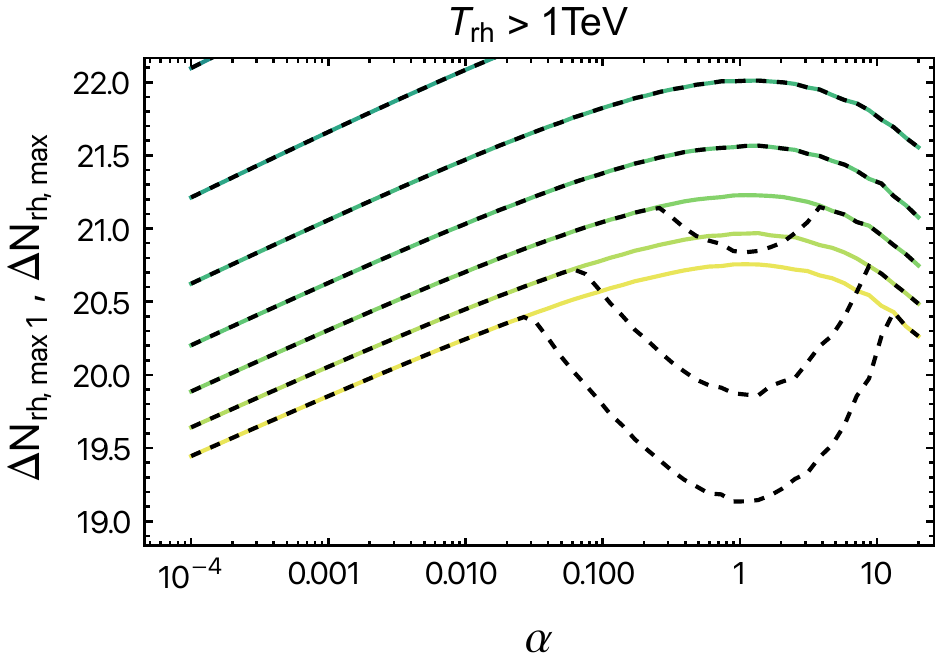}
    \end{subfigure}
    \caption{\textit{Left panel:} For T-models~\eqref{eq: T-model potential} with even $p\in [2,\,20]$ and $10^{-4}\leq \alpha \leq 20$ we represent $\Delta N_\text{rh, max 1}$, obtained from Eq.~\eqref{eq: first bound on duration of reheating}. 
    \textit{Right panel:} 
    For $p\geq 10$ we show $\Delta N_\text{rh, max}\equiv \min \left( \Delta N_\text{rh, max 1},\, \Delta N_\text{rh, max 2}\right)$ (dashed, black line), and compare it with $\Delta N_\text{rh, max 1}$ (colored lines).}
    \label{fig: maximum duration of reheating}
\end{figure}
In the left panel of Fig.~\ref{fig: maximum duration of reheating} we show $\Delta N_\text{rh, max 1}$  as a function of $\alpha$, where each line corresponds to fixed $p$. 
For $\rho_\text{rh, min 1}$ fixed by Eq.~\eqref{eq: first bound on duration of reheating}, larger $p$ values yield larger equation of state parameter~\eqref{eq: equation of state parameter as function of p}, thereby decreasing $\Delta N_\text{rh, max 1}$, see Eq.~\eqref{eq: Delta N rh def}. 
The dependence of $\Delta N_\text{rh, max 1}$ on $\alpha$ reflects the shape of $\rho_\text{end}(p,\,\alpha)$, and it is very mild due to $\Delta N_\text{rh, max 1}$ being sensitive to $\rho_\text{end}$ only logarithmically. 

As we shall demonstrate explicitly in Sec.~\ref{sec: inflationary predictions}, when $p\geq 6$ the longer reheating is the larger is $n_s$, bringing T-models closer to P-ACT-LB-BK18 observations.
In Sec.~\ref{sec: how far can we push ns in these models} our aim is to establish how close to unity $n_s$ can be within T-models. 
To this end, we should allow reheating to be as long as possible. 
While the constraint in Eq.~\eqref{eq: first bound on duration of reheating} is well justified by having to allow for baryogenesis after reheating, it can be relaxed to the minimum requirement of having reheating complete before BBN. 
In analogy with Eq.~\eqref{eq: first bound on duration of reheating}, this reads 
\begin{equation}
\label{eq: BBN bound on duration of reheating}
    \rho_\text{rh} \geq \rho_\text{rh,min 1} \equiv \frac{\pi^2}{30}\; g_{\star,0.1\, \text{MeV}} \; \left(0.1\,\text{MeV}\right)^4  \;,
\end{equation}
where $g_{\star,0.1\, \text{MeV}}=10.75$. 
\begin{figure}
\centering
\captionsetup[subfigure]{justification=centering}
    \begin{subfigure}{.48\textwidth}
        \includegraphics[width=\textwidth]{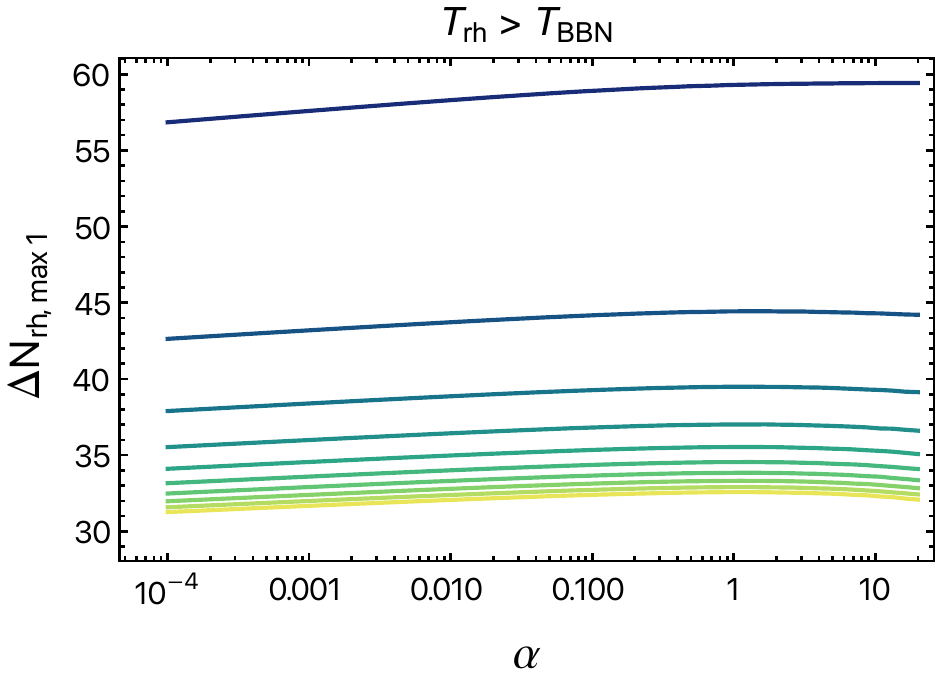}
    \end{subfigure}
    \begin{subfigure}{.48\textwidth}
        \includegraphics[width=\textwidth]{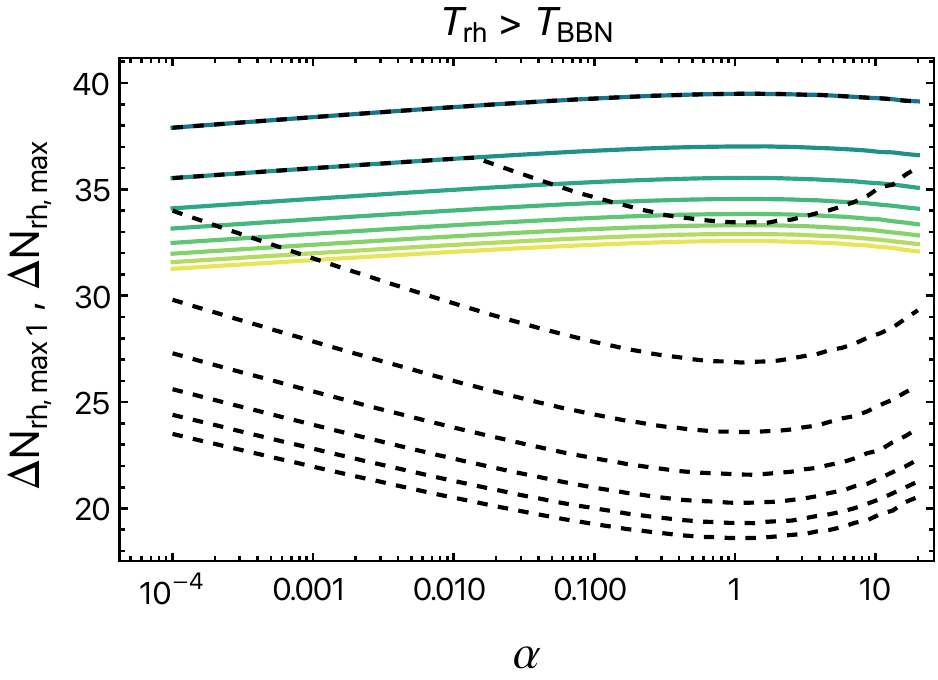}
    \end{subfigure}
    \begin{subfigure}{.6\textwidth}
        \includegraphics[width=\textwidth]{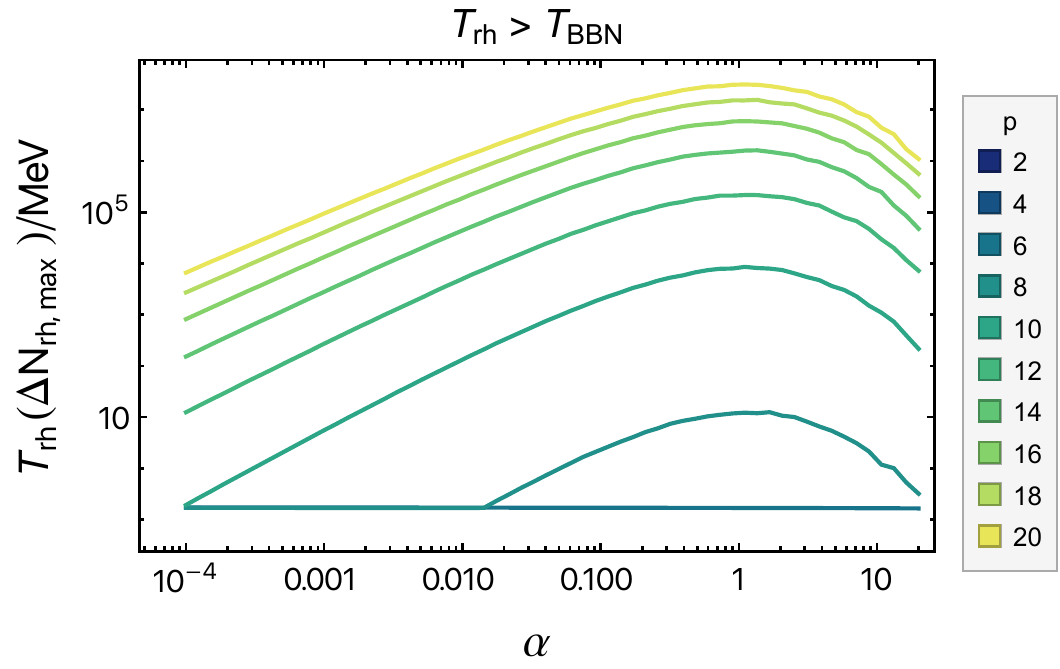}
    \end{subfigure}
    \caption{\textit{Top, left panel:} For T-models~\eqref{eq: T-model potential} with even $p\in [2,\,20]$ and $10^{-4}\leq \alpha \leq 20$ we represent $\Delta N_\text{rh, max 1}$, obtained from Eq.~\eqref{eq: BBN bound on duration of reheating}. 
    The color legend is the same as in the bottom panel. 
    \textit{Top, right panel:}  
    For $p\geq 6$ we represent $\Delta N_\text{rh, max}\equiv \min \left( \Delta N_\text{rh, max 1},\, \Delta N_\text{rh, max 2}\right)$ (dashed, black line), and compare it with $\Delta N_\text{rh, max 1}$ (colored lines).
    \textit{Bottom panel:} Minimum reheating temperature for models with $p\geq 6$, obtained by using $\Delta N_\text{rh, max}$. 
    For reference, the temperature at the time of BBN is $T_\text{BBN}=0.1 \, \text{MeV}$.} 
    \label{fig: BBN maximum duration of reheating}
\end{figure}
By following the same procedure described below Eq.~\eqref{eq: first bound on duration of reheating} we obtain $\Delta N_\text{rh, max 1}$ from Eq.~\eqref{eq: BBN bound on duration of reheating} and show the results in the top-left panel of Fig.~\ref{fig: BBN maximum duration of reheating}. 
Since now $T_\text{rh}$ is allowed to be much lower than in Eq.~\eqref{eq: first bound on duration of reheating}, $\Delta N_\text{rh, max 1}$ is much larger than the corresponding results in Fig.~\ref{fig: maximum duration of reheating}. 

\subsubsection*{$\bm{\Delta N_\text{eff}}$ bound for $\bm{p>4}$:  $\bm{\Delta N_\text{rh, max 2}}$}
For models with a stiff equation of state during reheating $(p>4 \to \bar w>1/3)$, we impose a second upper bound on the duration of reheating, which we label $\Delta N_\text{rh, max 2}$,
coming from the constraint on the energy density of primordial GWs~\cite{Caprini:2018mtu} 
\begin{equation}
\label{eq: Omega GW Neff bound}
    h^2 \int_{f_\text{rh}}^{f_\text{end}} \mathrm{d}\ln f \; \Omega_\text{GW}(f,\,\tau_0) < 5.6 \times 10^{-6} \, \Delta N_\text{eff} \;. 
\end{equation}
Here $H_0=100 h \,\text{km}/\text{s}/\text{Mpc}$, where we will take $h=0.67$~\cite{Planck:2018vyg}, and we use the bound on the effective number of relativistic degrees of freedom $\Delta N_\text{eff}<0.17$ $(95\%\, \text{C.L.})$ from the P-ACT-LB data combination~\cite{ACT:2025fju, ACT:2025tim}. 
In Eq.~\eqref{eq: Omega GW Neff bound} we have introduced the energy density in GWs, $\rho_\text{GW}$, per unit logarithmic (comoving) wavenumber, normalised with respect to the critical energy density, $\rho_\text{crit} \equiv 3 M_\text{Pl}^2 H^2$, 
\begin{equation}
\label{eq: Omega GW def}
    \Omega_\text{GW}(k,\,\tau) \equiv \frac{1}{\rho_\text{crit}} \frac{\mathrm{d}\rho_\text{GW}}{\mathrm{d}\ln k}=  \frac{1}{12} \frac{k^2}{a(\tau)^2 H(\tau)^2} \, T(k,\, \tau)\,  \mathcal{P}_t(k) \;.
\end{equation}
Here, $\tau$  is conformal time, defined by $\mathrm{d}\tau \equiv \mathrm{d}t/a(t)$, and we have explicitly separated the primordial signal, $\mathcal{P}_t(k)$, from the time- and scale-dependent transfer function, $T(k,\, \tau)$, that accounts for the evolution of GWs from
the moment they re-enter the horizon up to $\tau$, including the dependence on the equation of state parameter, $\bar w$.

Let us explain \emph{why} the GW bound in Eq.~\eqref{eq: Omega GW Neff bound} is particularly constraining for models with stiff reheating. 
During inflation, primordial GWs are produced from tensor fluctuations of the metric (see Sec.~\ref{sec: constraining inflation with CMB observations}), with power spectrum
\begin{equation}
    \label{eq: primordial tensor spectrum}
    \mathcal{P}_t (k) = \frac{2 H^2}{\pi^2 M_\text{Pl}^2}\Big|_{k=k_\text{CMB}} \left(\frac{k}{k_\text{CMB}} \right)^{-2\epsilon_1 +\mathcal{O}(\epsilon^2)}  \;,
\end{equation}
where we have pivoted the signal around the CMB comoving wavenumber, $k_\text{CMB}$, and included the slow-roll-suppressed spectral tilt. 
When $\bar w>1/3$, the approximately scale-invariant spectrum of primordial GWs modes that re-enter the horizon during reheating acquire a blue-tilted spectrum with spectral index $2(3\bar w-1)/(3\bar w+1)$~\cite{Figueroa:2019paj}, see Eq.~\eqref{eq: Omega GW for reheating}. 
The larger $p$ is then the larger the blue tilt, see Eq.~\eqref{eq: equation of state parameter as function of p}.
The modes involved are those that re-entered the horizon between the end of inflation and the end of reheating.  
For this reason, in Eq.~\eqref{eq: Omega GW Neff bound} the integral is performed over frequencies $f_\text{rh}\leq f\leq f_\text{end}$, where the extrema of this band are the frequencies of modes that re-entered the horizon at the end of reheating and at the end of inflation (as observed today, $f=(ck)/[2\pi a(t_0)]$).
Since modes that are blue-tilted and contribute to Eq.~\eqref{eq: Omega GW Neff bound} exited the horizon towards the end of inflation, we approximate Eq.~\eqref{eq: primordial tensor spectrum} with the scale-invariant quantity
\begin{equation}
    \label{eq: P_gamma end of inflation}
    \mathcal{P}_t \approx  \frac{2 H_\text{end}^2}{\pi^2 M_\text{Pl}^2} \;. 
\end{equation}
By modeling the transition from reheating into radiation domination as instantaneous, one obtains~\cite{Figueroa:2019paj}
\begin{equation}
\label{eq: Omega GW for reheating}
    \Omega_\text{GW}(f,\,\tau_0) =  2.94\times 10^{-7}\,\left(\frac{106.75}{g_{\star,T_\text{rh}}} \right)^{1/3} \frac{H_\text{end}^2}{M_\text{Pl}^2}\, \times \mathcal{W}\left(\frac{f}{f_\text{rh}}\right)\, \times  \mathcal{T}_s \, \times \left(\frac{f}{f_\text{rh}}\right)^{2 \frac{3\bar w -1}{3\bar w+1}} \;, 
\end{equation}
where $\tau_0$ indicates the present day and we have taken $g_{\star,T_\text{rh}}=g_{s,T_\text{rh}}$, since reheating is complete before the two deviate from each other. 
We substitute $g_{\star,T_\text{rh}} = 106.75$ when we require $T_\text{rh}>1\,\text{TeV}$ and $g_{\star,T_\text{rh}} = 10.75$ when we require $T_\text{rh}>T_\text{BBN}$.  
    \footnote{In principle we should set $g_{\star,T_\text{rh}}$ to the appropriate value corresponding to each value of $T_\text{rh}$. 
    However the change of $g_{\star,T_\text{rh}}$ from $106.75$ to $10.75$, for example, only amplifies the GW signal by a factor $\sim2$, and we deem this error to be acceptable. 
    In practice we will be particularly interested in the minimum allowed values for $T_\text{rh}$ (see Secs.~\ref{sec: inflationary predictions} and~\ref{sec: how far can we push ns in these models}), in which case $g_{\star,T_\text{rh}}$ is set to the correct value.}
See Eqs.~(3.21) and~(3.25) in Ref.~\cite{Figueroa:2019paj} for explicit expressions for the window function $\mathcal{W}(x)$ and the dimensionless constant $1<\mathcal{T}_s<4/\pi$ for $p>4$. 
In the limit $f\gg f_\text{rh}$, $\mathcal{W}(x)\to1$. 

Due to the shape of the signal~\eqref{eq: Omega GW for reheating}, the integral in Eq.~\eqref{eq: Omega GW Neff bound} can be approximated for $\bar{w}>1/3$ as 
\begin{equation}
\label{eq: approx Omega GW integral}
    \int_{f_\text{rh}}^{f_\text{end}} \mathrm{d}\ln f \; \Omega_\text{GW}(f,\,\tau_0) \approx 2.94\times 10^{-7} \,\left(\frac{106.75}{g_{\star,T_\text{rh}}} \right)^{1/3}   \,\frac{H_\text{end}^2}{M_\text{Pl}^2} \, \frac{3\bar w +1}{2 (3\bar w -1 )} \,\mathcal{T}_s \times \,\left(\frac{f_\text{end}}{f_\text{rh}}\right)^{2 \frac{3\bar w -1}{3\bar w +1 }} \;.
\end{equation} 
Primordial GWs enhanced due to stiff reheating contribute to the radiation energy budget, and potentially modify the tightly constrained expansion rate of the universe at the time of BBN. 
The longer reheating is, the larger is the separation between the frequencies $f_\text{rh}$ and $f_\text{end}$, and therefore the larger Eq.~\eqref{eq: approx Omega GW integral} is.
For this reason, the upper bound in Eq.~\eqref{eq: Omega GW Neff bound} can be translated in the upper limit $\Delta N_\text{rh}\leq \Delta N_\text{rh, max 2}$. 

For each model in our parameter space with $p\geq 6$, we first find $\Delta N_\text{rh, max 1}$ from Eq.~\eqref{eq: first bound on duration of reheating} (or~\eqref{eq: BBN bound on duration of reheating}), and compute the corresponding $H_\text{end}$, $T_\text{rh}$ and \begin{equation}
\label{eq:fendfrh}
    \frac{f_\text{end}}{f_\text{rh}} = \exp{\left(-\Delta N_\text{rh}\right)} \frac{H_\text{end}}{H_\text{rh}} = \exp\bigg(\Delta N_\text{rh}\frac{3\bar{w}+1}{2}\bigg) \;, 
\end{equation}
where $H_\text{rh} = \pi \,(90)^{-1/2} \, g_{\star, T_\text{rh}}^{1/2}\, T_\text{rh}^2/M_\text{Pl}$. 

We then substitute these in Eq.~\eqref{eq: approx Omega GW integral} together with Eq.~\eqref{eq: equation of state parameter as function of p} and check whether this satisfies the bound in Eq.~\eqref{eq: Omega GW Neff bound}.
If this is the case, the $\Delta N_\text{eff}$ bound does not produce a constraint on the duration of reheating stronger than the one from imposing Eq.~\eqref{eq: first bound on duration of reheating} (or~\eqref{eq: BBN bound on duration of reheating}). 
If Eq.~\eqref{eq: approx Omega GW integral} is larger than the right hand side of Eq.~\eqref{eq: Omega GW Neff bound}, we decrease progressively $\Delta N_\text{rh}$ (and consistently recompute $H_\text{end}$ and $f_\text{end}/f_\text{rh}$) until we find the first value that satisfies the bound~\eqref{eq: Omega GW Neff bound}, which we label $\Delta N_\text{rh, max 2}$. 
The results for $\Delta N_\text{rh, max}\equiv \min \left(\Delta N_\text{rh, max 1}, \,\Delta N_\text{rh, max 2}\right)$ are shown alongside those for $\Delta N_\text{rh, max 1}$ in the right panels in Fig.~\ref{fig: maximum duration of reheating} for $T_\text{rh}>1\,\text{TeV}$ and Fig.~\ref{fig: BBN maximum duration of reheating} for $T_\text{rh}>T_\text{BBN}$. 
The black-dashed line deviates from the corresponding colored one when $\Delta N_\text{rh, max 2}< \Delta N_\text{rh, max 1}$. 

When $T_\text{rh}>1\,\text{TeV}$, for $6 \leq p \leq 14$ Eq.~\eqref{eq: Omega GW Neff bound} does not provide a stronger constraint on reheating with respect to the requirement in Eq.~\eqref{eq: first bound on duration of reheating}, while for $p\geq 16$ the constraint from Eq.~\eqref{eq: Omega GW Neff bound} is stronger for a range of $\alpha$ values.
We have checked that the dependence of $\Delta N_\text{rh, max 2}$ on $\alpha$ for models with $\Delta N_\text{rh, max 2} < \Delta N_\text{rh, max 1}$ tracks the shape of $H_\text{end}(p,\,\alpha)$, which determines the baseline amplitude for the blue-tilted GWs. 
By translating the results for $\Delta N_\text{rh, max 2}$ into the corresponding minimum allowed temperature at the end of reheating, we find that for models with $\Delta N_\text{rh, max 2} < \Delta N_\text{rh, max 1}$ $T_\text{rh}$ is only slightly larger than $1\,\text{TeV}$, being at most $\sim8\,\text{TeV}$. 

When $T_\text{rh}>T_\text{BBN}$, reheating is allowed to last much longer, and therefore the constraint from Eq.~\eqref{eq: Omega GW Neff bound} is stronger than the one from Eq.~\eqref{eq: BBN bound on duration of reheating} for more models than in the case above. 
When $T_\text{rh}>T_\text{BBN}$, models with $p=6$ are not affected, but for $p=8$ ($p\geq 10$) $\Delta N_\text{rh, max 2}$ is smaller than $\Delta N_\text{rh, max 1}$ for a range of $\alpha$ values (the whole range of $\alpha$ considered).
We translate $\min \left(\Delta N_\text{rh, max 1}, \,\Delta N_\text{rh, max 2}\right)$ into the corresponding maximum temperature at the completion of reheating and plot the results in the bottom panel of Fig.~\ref{fig: BBN maximum duration of reheating}. 
For models affected by the GW constraint~\eqref{eq: Omega GW Neff bound}, $T_\text{rh}(\Delta N_\text{rh, max})$ is higher than $T_\text{BBN}$, and it gets lifted up to $\sim 10 \, \text{TeV}$ for a model with $p=20$ and $\alpha=1$. 

\section{Inflationary predictions}
\label{sec: inflationary predictions}
In this section we obtain the $n_s$ and $r$ inflationary predictions for T-models with even $p\in [2,\,20]$ and $10^{-4}\leq \alpha\leq 20$.
We choose to include large-$p$ models because they are characterised by a stiff equation-of-state during reheating which can lead to values for $n_s$ closer to unity.
We choose a uniform prior on $\log_{10}\alpha$, and bound $\alpha$ from below at $10^{-4}$~\cite{Iacconi:2023mnw}.
This is because $n_s$ is proportional to $\log_{10}\alpha$ due to the $\alpha$-dependence of $\Delta N_\text{CMB, inst rh}$, and therefore smaller $\alpha$ would lead to smaller $n_s$. 

We first discuss how the $n_s$ and $r$ inflationary predictions are computed in Sec.~\ref{sec: how ns and r are computed}, then present a selection of our results for $T_\text{rh}>1\,\text{TeV}$ and $T_\text{rh}>T_\text{BBN}$ in Sec.~\ref{sec: predictions vs PACT}. 
In Sec.~\ref{sec: Parameter space constraints from P-ACT-LB-BK18} we discuss current bounds to the T-models parameter space in light of P-ACT-LB-BK18 data.  

\subsection[\texorpdfstring{Method for obtaining ${n_s}$ and ${r}$}{Method for obtaining ns and r}]{Method for obtaining $\bm{n_s}$ and $\bm{r}$}
\label{sec: how ns and r are computed}
To compute $n_s$ and $r$, we employ the analytical expressions obtained at first-order in the slow-roll expansion~\cite{Kallosh:2013yoa, Kallosh:2013hoa}
\begin{align}
\label{eq: ns analytic}
    n_s& \approx 1-\frac{2\Delta N_\text{CMB}+\frac{1}{p}\sqrt{3\alpha(3\alpha+p^2)}+\frac{3\alpha}{2}}{\Delta N_\text{CMB}^2 +\frac{\Delta N_\text{CMB}}{p}\sqrt{3\alpha(3\alpha+p^2)}+\frac{3\alpha}{4}} \;, \\
\label{eq: r analytic}
    r& \approx \frac{12\alpha}{\Delta N_\text{CMB}^2 +\frac{\Delta N_\text{CMB}}{p}\sqrt{3\alpha(3\alpha+p^2)}+\frac{3\alpha}{4}} \;. 
\end{align}
These are given in terms of three parameters: $\Delta N_\text{CMB}$ and the potential parameters $p$ and $\alpha$. 
Rather than being a free parameter, $\Delta N_\text{CMB}$ is determined by the inflaton potential (and therefore depends itself on $p$ and $\alpha$), and by reheating's equation-of-state and duration, see Eq.~\eqref{eq: DN CMB}. 
The $p$-dependent terms in Eqs.~\eqref{eq: ns analytic} and~\eqref{eq: r analytic} are sizeable when $\alpha\gtrsim \mathcal{O}(1)$. 
Otherwise, for fixed reheating, $n_s$ and $r$ are insensitive to $p$.
    \footnote{This is because also $\Delta N_\text{CMB, inst rh}$ does not depend on $p$ when $\alpha \lesssim \mathcal{O}(1)$.}

Cosmological $\alpha$-attractors are often characterised by \emph{universal} predictions for $n_s$ and $r$. 
For T-models these are obtained from Eqs.~\eqref{eq: ns analytic} and~\eqref{eq: r analytic} in the large $\Delta N_\text{CMB}$ limit, yielding~\cite{Kallosh:2013hoa, Kallosh:2013yoa}
\begin{align}
\label{eq: ns universal}
    n_s & \approx 1-\frac{2}{\Delta N_\text{CMB}}\;, \\
\label{eq: r universal}
    r & \approx \frac{12\alpha}{{\Delta N_\text{CMB}}^2}  \;.
\end{align}
These are referred to as universal since in the limit $\Delta N_\text{CMB}\gg p,\, \alpha$ the predictions do not explicitly depend on the potential parameter $p$.
Note that Eqs.~\eqref{eq: ns universal} and~\eqref{eq: r universal} describe the universal predictions for all $\alpha$-attractors with exponential approach to the large-field potential plateau, e.g. also for the so-called E-models~\cite{Kallosh:2015zsa}. 

To obtain $n_s$ and $r$ from Eqs.~\eqref{eq: ns analytic} and~\eqref{eq: r analytic} we first need to compute $\Delta N_\text{CMB}$. 
For all models in the parameter space we iteratively solve Eq.~\eqref{eq: DN CMB} with $\Delta N_\text{rh}=0$ and for values of $V_0$ compatible with CMB observations to find $\Delta N_\text{CMB, inst rh}$, i.e. the component of $\Delta N_\text{CMB}$ that is completely determined by the inflaton potential. 
Then, we include the effect of reheating by producing a range of $\Delta N_\text{CMB}$ values from Eq.~\eqref{eq: DN CMB} with equation-of-state determined by Eq.~\eqref{eq: equation of state parameter as function of p} and $0\leq \Delta N_\text{rh}\leq \Delta N_\text{rh, max}$, where we use the results of Sec.~\ref{sec: duration of reheating} and define $\Delta N_\text{rh, max} \equiv \min \left(\Delta N_\text{rh, max 1}, \, \Delta N_\text{rh, max 2} \right)$. 
For each model, identified by the three parameters $\{p,\, \alpha, \Delta N_\text{rh}\}$, we obtain $n_s$ and $r$ by substituting $\Delta N_\text{CMB}$, $p$ and $\alpha$ in Eqs.~\eqref{eq: ns analytic} and~\eqref{eq: r analytic}. 

Before moving on to discuss our results, let us comment on the precision of our predictions. 
To this end, we focus on the scalar spectral tilt, as it is of foremost importance that this is determined precisely in order to compare inflationary predictions against CMB data.   
For a range of test models, we compute $n_s$ by substituting the numerical background solution and $\Delta N_\text{CMB}$ into the corresponding expression at second-order in the slow-roll expansion, and label the result as $n_{s,\, \text{so}}$. 
We then compare the theoretical error made in using Eq.~\eqref{eq: ns analytic} instead of the above, i.e. $|n_{s,\,\text{so}}-n_{s,\,\text{analytical}}|$, to the experimental error from the P-ACT-LB-BK18 analysis, $\sigma(n_s)$ in Eq.~\eqref{eq:ns P-ACT-LB-BK18}. 
More details are given in Appendix~\ref{app: theoretical error analytical prediction}, and results are shown in Fig.~\ref{fig:theoretical ns error first vs second order slow roll}. 
We find that $|n_{s,\,\text{so}}-n_{s,\,\text{analytical}}|$ is always at least one order of magnitude smaller than $\sigma(n_s)$. 
This result ensures that Eq.~\eqref{eq: ns analytic} is precise enough to be used when comparing T-models with P-ACT-LB-BK18 data.

\subsection[\texorpdfstring{Predictions for ${n_s}$ and ${r}$ confront P-ACT-LB-BK18}{Predictions for ns and r confront P-ACT-LB-BK18}]{Predictions for $\bm{n_s}$ and $\bm{r}$ confront P-ACT-LB-BK18}
\label{sec: predictions vs PACT}
\begin{figure}
\centering
\captionsetup[subfigure]{justification=centering}
    \begin{subfigure}{.48\textwidth}
        \includegraphics[width=\textwidth]{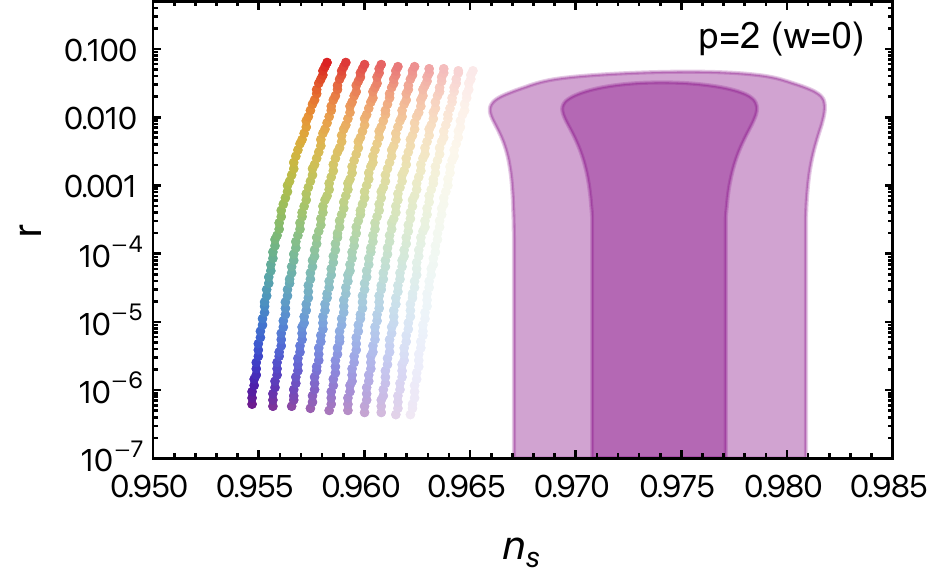}
    \end{subfigure}
    \begin{subfigure}{.48\textwidth}
        \includegraphics[width=\textwidth]{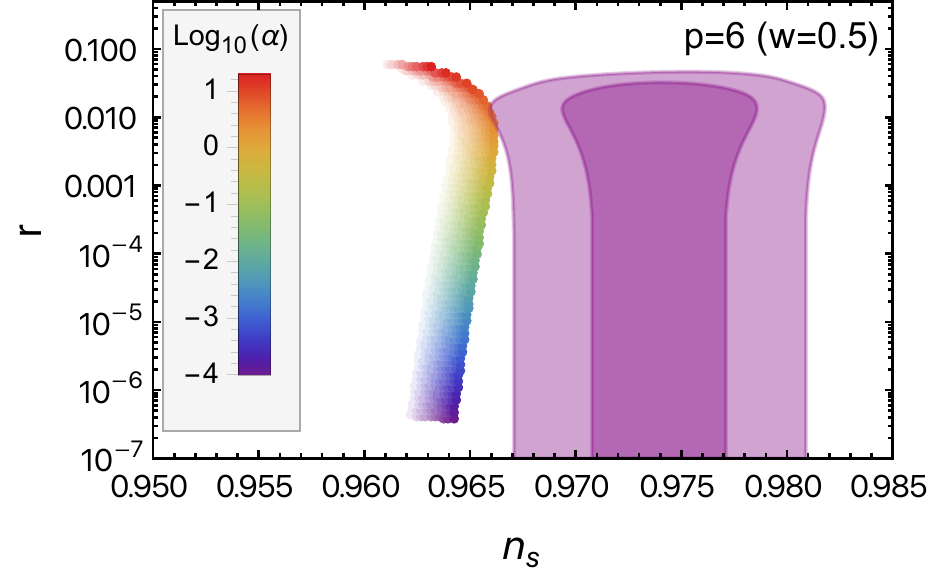}
    \end{subfigure}
    \begin{subfigure}{.48\textwidth}
        \includegraphics[width=\textwidth]{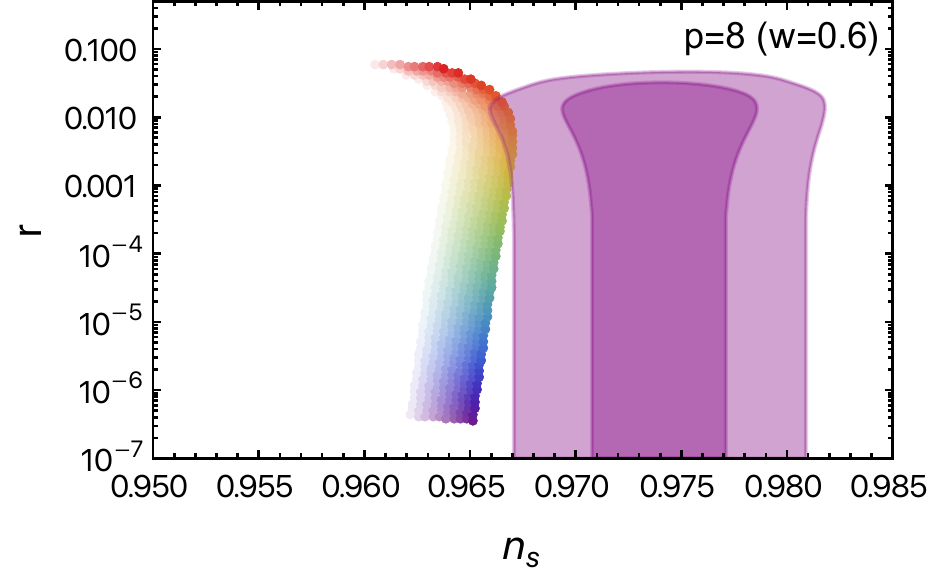}
    \end{subfigure}
    \begin{subfigure}{.48\textwidth}
        \includegraphics[width=\textwidth]{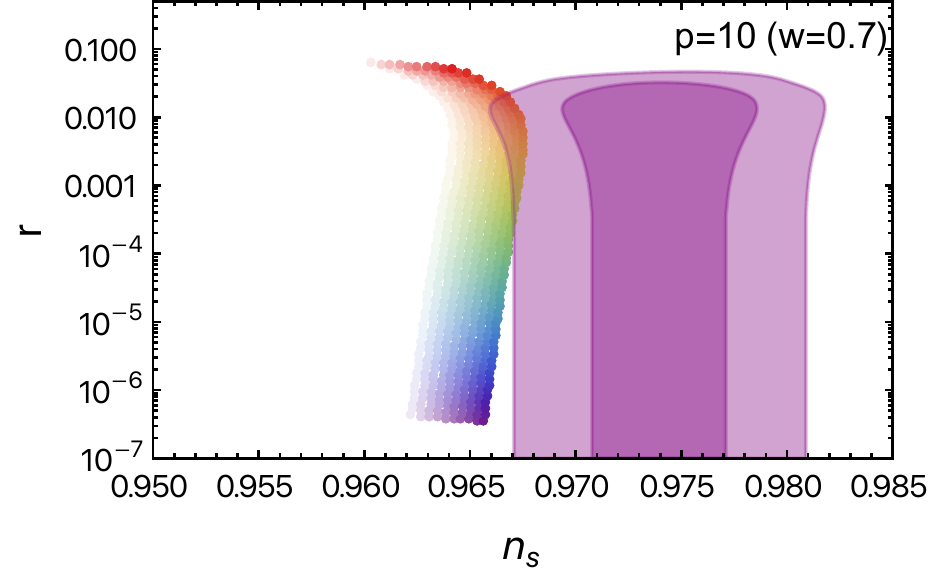}
    \end{subfigure}
\caption{Predictions in the $(n_s,\,r)$ plane for T-models with potential~\eqref{eq: T-model potential}. 
Each panel contains results for a different value of $p$ (see the top-right label).  
For fixed $p$ we produce predictions for $10^{-4}\leq \alpha\leq 20$ (see the top-right panel for the corresponding color legend). 
For each $(p,\,\alpha)$ pair, we consider 10 reheating scenarios with $0\leq \Delta N_\text{rh}\leq \min\left(\Delta N_\text{rh, max 1}, \, \Delta N_\text{rh, max 2}\right)$, where we have computed $\Delta N_\text{rh, max 1}$ by requiring that reheating is complete before the universe reaches $1\,\text{TeV}$. 
We distinguish predictions for different $\Delta N_\text{rh}$ by using a different shading of the $\alpha$-identifying color. 
Lighter (darker) shading indicates a shorter (longer) reheating stage. 
The pink contours represent the $68\%\,\text{C.L.}$ and $95\%\,\text{C.L.}$ P-ACT-LB-BK18 posterior in the $(n_s,\,r)$ plane~\cite{ACT:2025tim}. 
}
     \label{fig:predictions in ns and r plane}
\end{figure}
We display in Fig.~\ref{fig:predictions in ns and r plane} results in the $(n_s,\, r)$ plane obtained for $T_\text{rh}>1\,\text{TeV}$. 
For the models represented here, the GW constraint~\eqref{eq: Omega GW Neff bound} is always weaker than the one in Eq.~\eqref{eq: first bound on duration of reheating}, and therefore $\Delta N_\text{rh, max} = \Delta N_\text{rh, max 1}$. 
For $p=2$ and fixed $\alpha$, an extended reheating stage decreases the value of $n_s$, increasing the tension of this region of the parameter space with current P-ACT-LB-BK18 data. 
We have checked that models with $p=4$ are also not compatible with the aforementioned data. 
In this case $n_s$ and $r$ are insensitive to details of reheating, as $p=4$ corresponds to $w=1/3$, see Eq.~\eqref{eq: DN CMB}. 
When $p>4$ reheating contributes in the opposite way, and makes $n_s$ larger as $\Delta N_\text{rh}$ increases.
For $p=6$ the predictions barely enter the $95\%$ C.L. contour when reheating is at its maximum allowed duration, while more and more models are compatible at least at $2\sigma$ when $p=8$ and $p=10$. 
The fact that the number of compatible models increases when going from $p=8$ to $p=10$ shows that while on one hand $\Delta N_\text{rh, max 1}$ decreases the larger $p$ is (see Fig.~\ref{fig: maximum duration of reheating}), the $\bar w$-dependent multiplying factor in Eq.~\eqref{eq: DN CMB}, $-(1-3\bar w)/4$, amplifies the effect of reheating, and determines the overall $n_s$ behavior. 

When $p>2$ and $\alpha\gtrsim \mathcal{O}(1)$ the predictions show a turn-around in the $(n_s,\, r)$ plane, which means that (for sufficiently large $p$ value) they enter the P-ACT-LB-BK18 contours only for a finite range of $\alpha$ values. 
We will show how this can be used to constrain $\alpha$ in Sec.~\ref{sec: Parameter space constraints from P-ACT-LB-BK18}. 
Note that one would erroneously miss this important effect if the universal predictions~\eqref{eq: ns universal} and~\eqref{eq: r universal} were used instead of Eqs.~\eqref{eq: ns analytic} and~\eqref{eq: r analytic}. 

\begin{figure}
\centering
\captionsetup[subfigure]{justification=centering}
    \begin{subfigure}{.48\textwidth}
        \includegraphics[width=\textwidth]{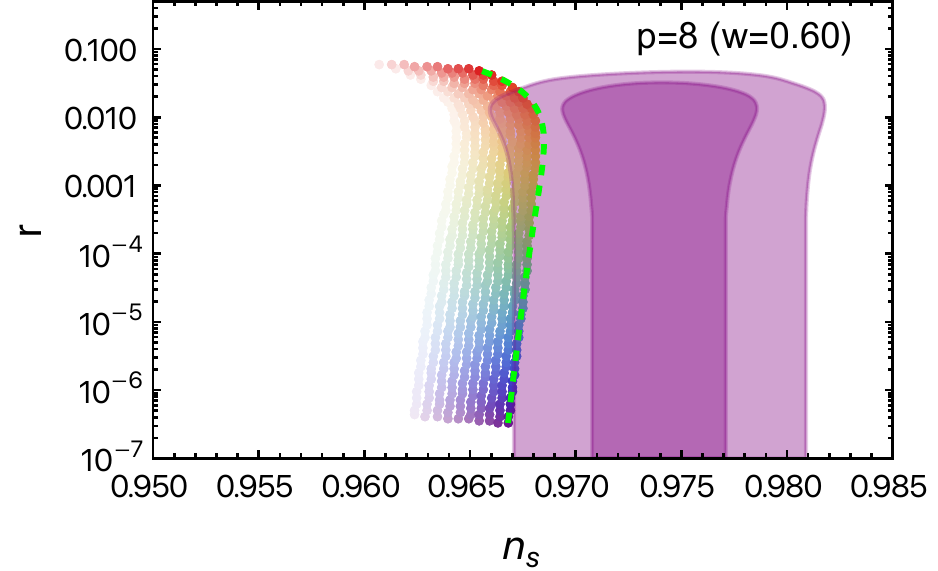}
    \end{subfigure}
    \begin{subfigure}{.48\textwidth}
        \includegraphics[width=\textwidth]{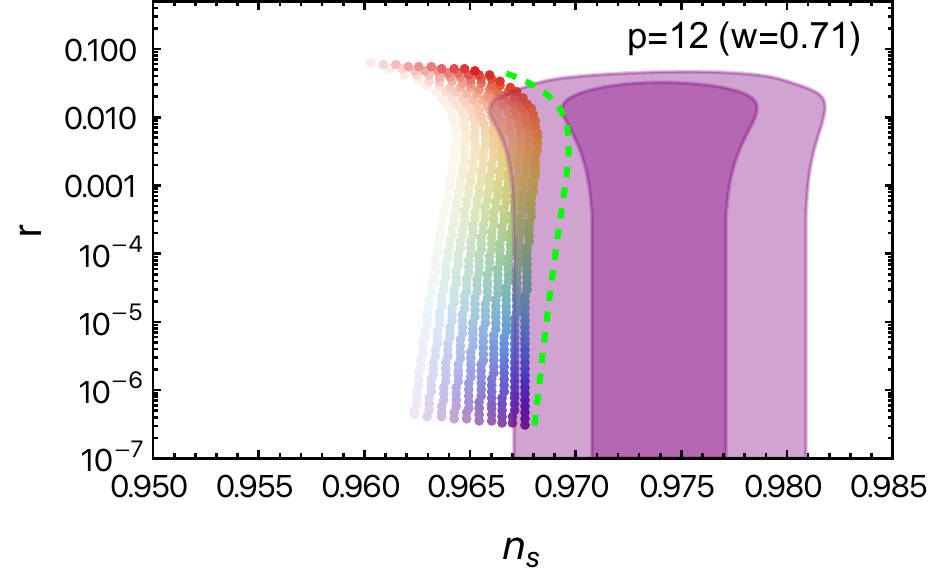}
    \end{subfigure}
    \begin{subfigure}{.48\textwidth}
        \includegraphics[width=\textwidth]{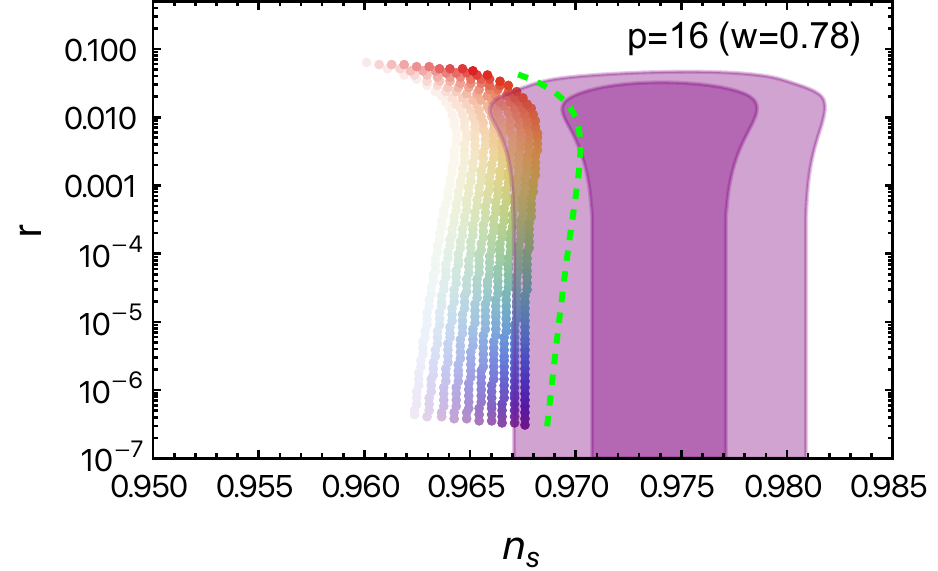}
    \end{subfigure}
    \begin{subfigure}{.48\textwidth}
        \includegraphics[width=\textwidth]{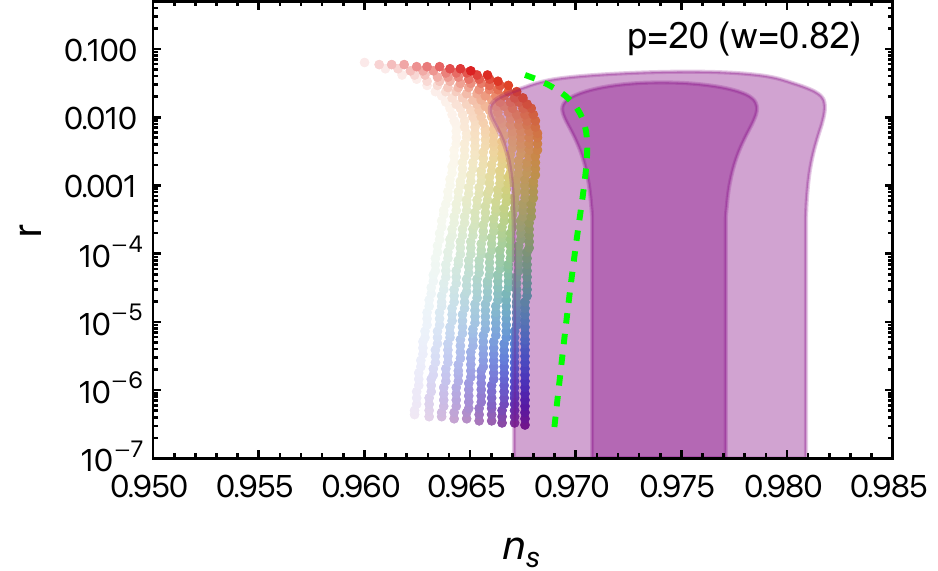}
    \end{subfigure}
\caption{Same as in Fig.~\ref{fig:predictions in ns and r plane}, but here we focus on large $p$ values and allow the temperature at the end of reheating to be as low as $T_\text{BBN}$.  
The dashed, green line indicates the $(n_s,\,r)$ predictions for $\Delta N_\text{rh}=\Delta N_\text{rh, max 1}$, i.e. computed without taking into account the GW constraint in Eq.~\eqref{eq: Omega GW Neff bound}. 
}
     \label{fig: BBN predictions in ns and r plane}
\end{figure}
In Fig.~\ref{fig: BBN predictions in ns and r plane} we display the results obtained by requiring that $T_\text{rh}>T_\text{BBN}$, and sample the large-$p$ region of the parameter space evenly.
By comparing the $p=8$ panels in Figs.~\ref{fig:predictions in ns and r plane} and~\ref{fig: BBN predictions in ns and r plane} one sees that allowing longer reheating (i.e. smaller $T_\text{rh}$) increases the compatibility with P-ACT-LB-BK18 data. 
When $p=\{12,\, 16,\, 18\}$, the GW constraint~\eqref{eq: BBN bound on duration of reheating} is stronger than the one in Eq.~\eqref{eq: BBN bound on duration of reheating}, see Fig.~\ref{fig: BBN maximum duration of reheating}. 
We illustrate the importance of including the GW constraint by showing the predictions obtained with the maximum allowed duration of reheating, $\Delta N_\text{rh}=\Delta N_\text{rh, max 2}$, against those that one would have obtained instead by using Eq.~\eqref{eq: BBN bound on duration of reheating} only.   
Note that the two aforementioned lines of $(n_s,\, r)$ predictions are not parallel, due to the non-trivial $p$ and $\alpha$ dependence of $\Delta N_\text{rh, max 2}$, see Fig.~\ref{fig: BBN maximum duration of reheating}. 

Due to the asymptotic behavior of $\bar w$~\eqref{eq: equation of state parameter as function of p} for large $p$ and (to a lesser extent) to the GW constraint~\eqref{eq: Omega GW Neff bound}, there is no appreciable increase in the P-ACT-LB-BK18 $95\%$ C.L. region covered by compatible models when moving from $p=16$ to $p=20$. 
This shows that increasing $p$ does not always ameliorate the compatibility with P-ACT-LB-BK18, but, rather, we expect it to saturate.  
This means that if future CMB measurements will support the trend towards larger $n_s$, T-models might be excluded altogether. 
Pushing T-models $n_s$ predictions to their limit will be the subject of Sec.~\ref{sec: how far can we push ns in these models}. 

\subsection{Parameter space constraints from P-ACT-LB-BK18}
\label{sec: Parameter space constraints from P-ACT-LB-BK18}

In order to obtain constraints on the inflationary parameters $\{p,\, \alpha, \Delta N_\text{rh}\}$, one would have to use a discrete prior for $p$, impose a prior on $\Delta N_\text{rh}$ over the (model-dependent) range $0\leq \Delta N_\text{rh}\leq \Delta N_\text{rh, max}(p,\,\alpha)\equiv \min\left(\Delta N_\text{rh, max 1}(p,\,\alpha), \,\Delta N_\text{rh, max 2}(p,\,\alpha) \right)$, and perform a Bayesian analysis to compare inflationary predictions against (at a minimum) CMB data. 
We did a similar analysis in Ref.~\cite{Iacconi:2023mnw} for fixed $p=2$ and $p=4$ to constrain $\Delta N_\text{rh}$ and $\log_{10}\alpha$ with \textit{Planck} and BK18 data. 
We do not pursue this avenue here with the new ACT data~\cite{ACT:2025fju}. 
First, this is because ACT data and DESI BAO data are in tension under the assumption of the standard cosmological model~\cite{Ferreira:2025lrd}, and this should be resolved before using ACT data to meaningfully constrain inflation. 
Second, our aim is a theoretical exploration of the large $p$ (and therefore large $n_s$) regime of T-models.

Nevertheless, it is useful to explore what CMB data can tell us about T-models \emph{if} --pending tensions being resolved-- future CMB measurements will support the preference seen with ACT data for $n_s$ larger than that determined by \textit{Planck}. 
To this end, we employ the 2-dimensional P-ACT-LB-BK18 posterior in the $(n_s,\,r)$ plane to illustrate how future observations might constrain the $(\alpha,\,p)$ parameter space, with $p\geq 6$. 
For each model, we compute the $n_s$ and $r$ predictions by fixing the reheating scenario that maximises $n_s$, i.e. $\Delta N_\text{rh}=\Delta N_\text{rh, max}$.
\footnote{Note that $r$ depends on the duration of reheating only mildly, see Figs.~\ref{fig:predictions in ns and r plane} and~\ref{fig: BBN predictions in ns and r plane}.}
This allows us to obtain the \emph{weakest} constraints in the parameter space from P-ACT-LB-BK18 data. 
\begin{figure}
\centering
\captionsetup[subfigure]{justification=centering}
    \begin{subfigure}{.6\textwidth}
        \includegraphics[width=\textwidth]{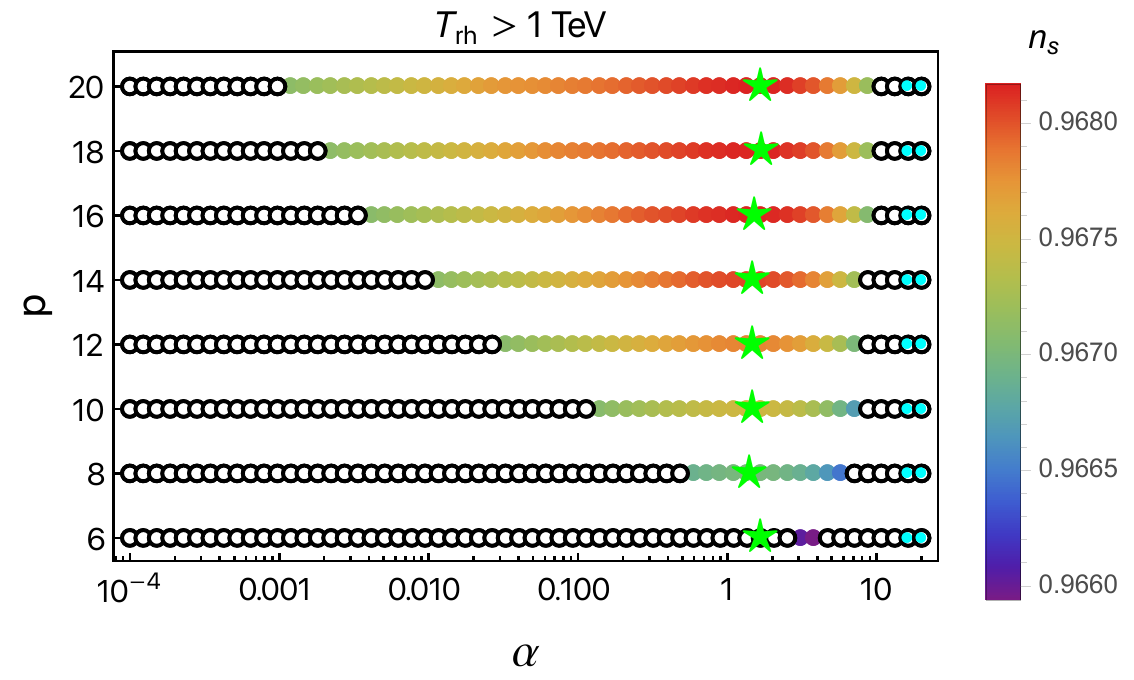}
    \end{subfigure}
    \begin{subfigure}{.6\textwidth}
        \includegraphics[width=\textwidth]{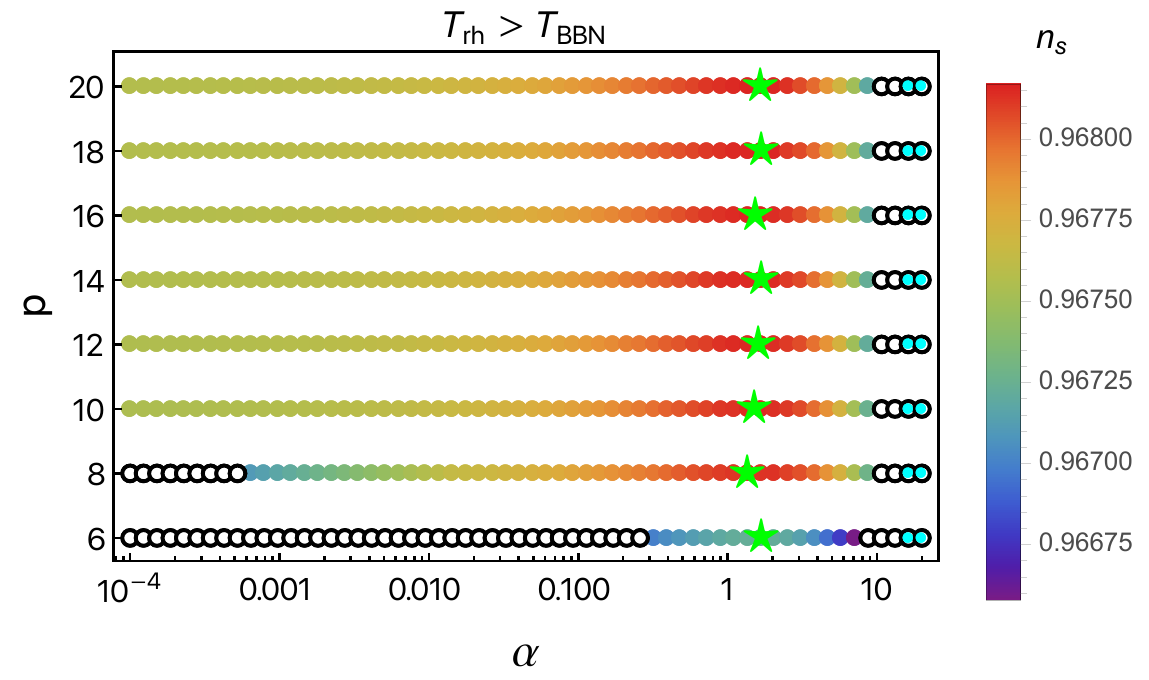}
    \end{subfigure}
    \caption{Values of the scalar spectral tilt, $n_s$, computed using Eq.~\eqref{eq: ns analytic} over the $(\alpha, \,p)$ parameter space.
    For each model we fix the reheating scenario that maximises $n_s$, see main text for more details.
    The top (bottom) panel contains results obtained by requiring $T_\text{rh}>1\,\text{TeV}$ ($T_\text{rh}>T_\text{BBN}$). 
    Models which are not compatible with the P-ACT-LB-BK18 (2-dimensional) $95\% \, \text{C.L.}$ constraints on $(n_s,\,r)$ are marked with an empty circle. 
    Models excluded by the $95\% \, \text{C.L.}$ upper limit on $r$ only, see Eq.~\eqref{eq:r P-ACT-LB-BK18}, are colored in cyan. 
    For each $p$ value, a green star indicates the value of $\alpha$ that maximises $n_s$.
    Note that we choose to represent each model with a point, rather than by using a continuous map, to emphasise that $p$ has a discrete prior.}
    \label{fig: ACT constraints on parameter space}
\end{figure}
We represent in Fig.~\ref{fig: ACT constraints on parameter space} values of $n_s$ in the $(\alpha, \, p)$ parameter space, together with the $95\% \, \text{C.L.}$ constraints obtained from $n_s$ \emph{and} $r$ jointly, and $r$ only.

When $T_\text{rh}>1\,\text{TeV}$, for each $p$, P-ACT-LB-BK18 data are compatible with only a range of $\alpha$ values, which is very small for $p=6$, and increases for larger $p$.
For fixed $p$, the range of allowed $\alpha$ values depends on the reheating scenario, and the constraint on $\alpha$ weakens when we allow $T_\text{rh}>T_\text{BBN}$.
In this case, as soon as $p\geq 10$ the constraint on $\alpha$ is limited by the $\alpha$ prior. 

It is well known that CMB constraints on the tensor-to-scalar ratio can be used to place an upper bound on the $\alpha$ parameter, see Eq.~\eqref{eq: r universal}. 
We pointed out in Ref.~\cite{Iacconi:2023mnw} that including the $\alpha$-dependence of $\Delta N_\text{CMB, inst rh}$ allows one to constrain the magnitude of $\alpha$ from below by using \textit{Planck} and BK18 data, in particular the $n_s$ measurement. 
Here, we see a different way in which a precise $n_s$ measurement can play out in constraining $\alpha$. 
By comparing the constraints obtained from $n_s$ and $r$ jointly against those obtained from $r$ only, one sees in Fig.~\ref{fig: ACT constraints on parameter space} an upper bound on $\alpha$ that is stronger than the one obtained from $r$ only.
As in Ref.~\cite{Iacconi:2023mnw}, the lower limit on $\alpha$ is due to the logarithmic dependence of $n_s$ on $\alpha$, while the upper bound is due to the $p$-dependence of Eq.~\eqref{eq: ns analytic} when $\alpha\gtrsim \mathcal{O}(1)$. 
These two effects can be easily visualised in Fig.~\ref{fig:predictions in ns and r plane} by comparing the $(n_s,\,r)$ predictions for, e.g., $p=8$ and fixed $\Delta N_\text{rh}=\Delta N_\text{rh, max}$ against the $2\sigma$ contours. 

For each $p$ we find that $n_s$ is maximised by $\alpha\approx 1$. 
Interestingly, seven models with $\alpha=\mathcal{O}(1)$, known as Poincar\' e models $(3\alpha = \{1, \,2,\, 3,\, 4,\,5, \, 6, \, 7\})$, have a special status within cosmological $\alpha$-attractors as their parent supergravity theory is characterised by additional symmetries~\cite{Ferrara:2016fwe, Kallosh:2017ced}. 
If CMB experiments confirm a preference for large $n_s$, these models are very well placed to deliver large $n_s$. 

\section[\texorpdfstring{How large can $\bm{n_s}$ be within T-models?}{How large can ns be within T-models?}]{How large can $\bm{n_s}$ be within T-models?}
\label{sec: how far can we push ns in these models}

ACT CMB data measure a primordial scalar spectral tilt larger than that determined by \textit{Planck}, see Eq.~\eqref{eq:ns P-ACT-LB-BK18}.
If this preference for values of $n_s$ closer to unity will be confirmed once the tension between ACT data and DESI BAO data is resolved and/or by future CMB data, what $n_s$ measurement would allow us to rule out T-models? 
Thanks to the lessons learned in Secs.~\ref{sec: inflation and reheating} and~\ref{sec: inflationary predictions}, in this section we will provide an answer to this question by computing the largest $n_s$ that T-models can produce.  
To this end, we will (i) consider the large-$p$ limit, (ii) fix $\alpha=1$ (see Fig.~\ref{fig: ACT constraints on parameter space}), and (iii) allow reheating to last as long as possible, by using the minimum requirement $T_\text{rh}>T_\text{BBN}$, while still imposing the GW constraint~\eqref{eq: Omega GW Neff bound}. 

Let us start exploring the large-$p$ limit by working in the extended range $p\in [8,\,40]$. 
When $\alpha=1$ and $p\geq 8$, the GW constraint on reheating~\eqref{eq: Omega GW Neff bound} is always stronger than the minimum requirement~\eqref{eq: BBN bound on duration of reheating} (see Fig.~\ref{fig: BBN maximum duration of reheating} for models up to $p=20$), and therefore $\Delta N_\text{rh, max}=\Delta N_\text{rh, max 2}$. 
\begin{figure}
\centering
\captionsetup[subfigure]{justification=centering}
    \begin{subfigure}{.48\textwidth}
        \includegraphics[width=\textwidth]{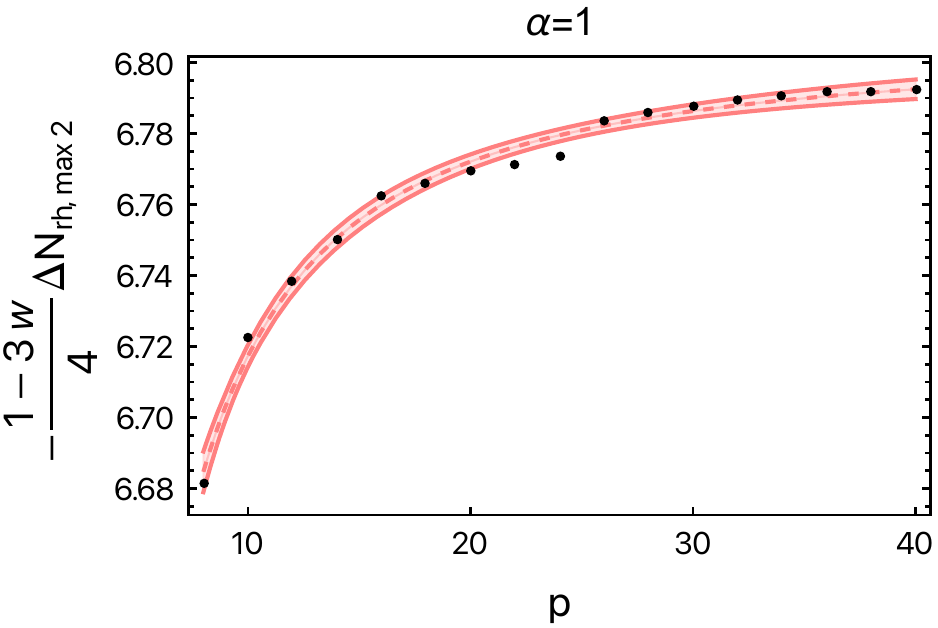}
    \end{subfigure}
    \begin{subfigure}{.48\textwidth}
        \includegraphics[width=\textwidth]{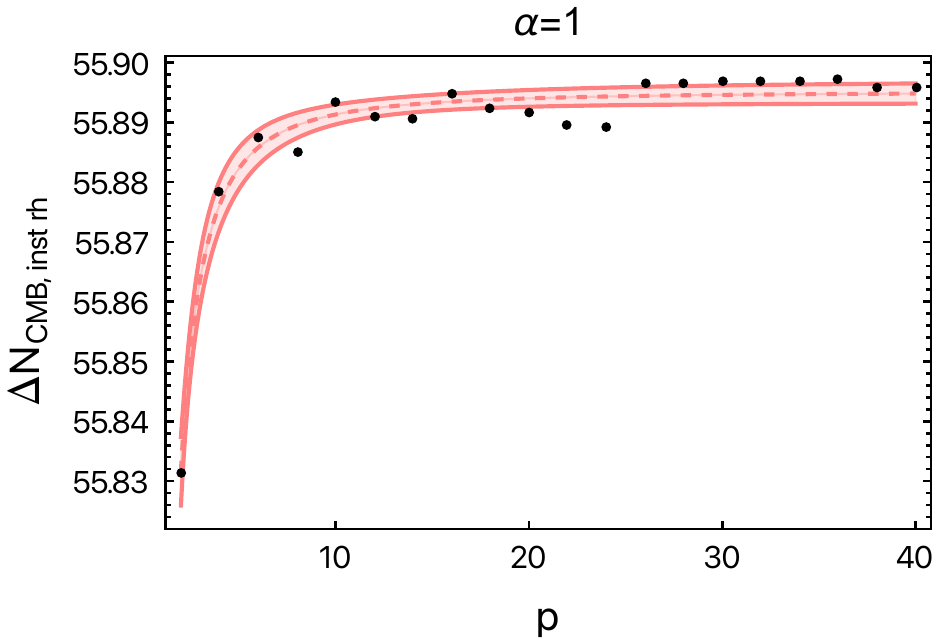}
    \end{subfigure}
\caption{\textit{Left panel:} Numerical values for the reheating contribution to $\Delta N_\text{CMB}$, $-(1-3\bar w)/{4}\,\Delta N_\text{rh, max 2}$, computed for $8\leq p\leq 40$. 
Here, $\Delta N_\text{rh, max 2}$ is the longest duration of reheating allowed such that $T_\text{rh}>T_\text{BBN}$ and reheating is compatible with the GWs constraint in Eq.~\eqref{eq: Omega GW Neff bound}. 
Numerical data points (black) are represented together with the fitting function~\eqref{eq: rh fit as function of p} (pink, dashed line), and the $95\%$ C.L. bands for the fit. 
\textit{Right panel:} Numerical values for $\Delta N_\text{CMB,inst rh}$ computed for $2\leq p\leq 40$. 
Numerical data points (black) are represented together with the fitting function~\eqref{eq: DN CMB inst rh fit as function of p} (pink, dashed line), and the $95\%$ C.L. bands for the fit. 
In both panels, we have fixed $\alpha=1$.}
     \label{fig: DNrh and DNCMB for p large}
\end{figure}
By following the same procedure outlined in Secs.~\ref{sec: inflation and reheating} and~\ref{sec: inflationary predictions}, we numerically compute $\Delta N_\text{CMB, inst rh}$ and the reheating contribution to $\Delta N_\text{CMB}$, that is $-(1-3\bar w)/{4} \,\Delta N_\text{rh, max 2}$, in Eq.~\eqref{eq: DN CMB}.
The numerical results are displayed in Fig.~\ref{fig: DNrh and DNCMB for p large} together with the fitting functions
    \footnote{Note that $\Delta N_\text{CMB, inst rh}$ approaches a constant value as soon as $p\gtrsim 6$, with the numerical results being affected by noise at the $\lesssim 0.1\%$ level for larger $p$. 
    Due to this noise in the numerical results, the quality of the fit improves by including also the results produced for $2\leq p< 8$.
    This is the reason why we have computed $\Delta N_\text{CMB, inst rh}$ over the range $p\in[2,\,40]$ instead of $p\in[8,\,40]$.}
\begin{align}
\label{eq: rh fit as function of p}
    -\frac{1-3\bar w}{4}\Delta N_\text{rh, max 2} &= 6.80455 -  2.32093 \, p^{-1.42508} \;, \\
\label{eq: DN CMB inst rh fit as function of p}
    \Delta N_\text{CMB, inst rh} & = 55.8952 - 0.212386 \, p^{-1.74213}  \;,  
\end{align}
which will allow us to extrapolate these quantities to even larger $p$.
Our results show that $\Delta N_\text{CMB, inst rh}$ depends on $p$ only very mildly, and it plateaus to $\sim 55.89$ as soon as $p\gtrsim 10$.  
Interestingly, the reheating contribution to $\Delta N_\text{CMB}$, shown in the left-hand panel in Fig.~\ref{fig: DNrh and DNCMB for p large}, saturates to a constant value. 
This is due to the asymptotic behavior of $\bar w$ and $\Delta N_\text{rh, max 2}$ in the large-$p$ limit, see Eq.~\eqref{eq: equation of state parameter as function of p} and Fig.~\ref{fig: BBN maximum duration of reheating}.  
This already indicates that there is a limit to the ability of an extended reheating stage with stiff equation of state to push $n_s$ to large values.
Due the asymptotic behavior of $-(1-3\bar w)/{4}\, \Delta N_\text{rh, max 2}$, we will be able to establish the \emph{largest} $n_s$ that T-models can produce.   

By substituting the numerical results for $\Delta N_\text{CMB, inst rh}$ and $-(1-3\bar w)/{4}\, \Delta N_\text{rh, max 2}$ in Eq.~\eqref{eq: DN CMB} and by using Eq.~\eqref{eq: ns analytic}, we obtain the models predictions for $n_s$. 
\begin{figure}
    \centering
    \includegraphics[width = 0.5\textwidth]{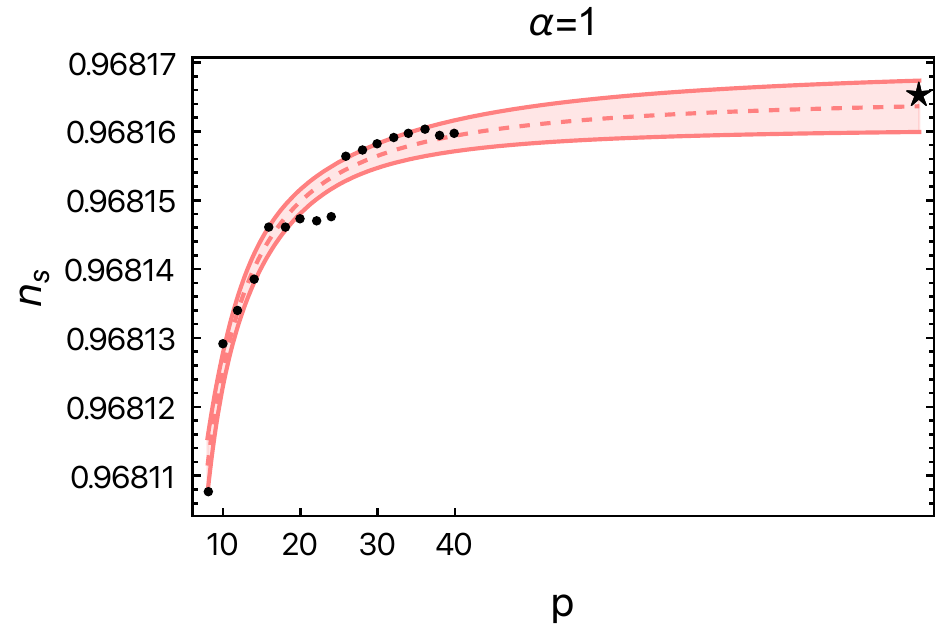}
    \caption{For $\alpha=1$ and $8\leq p\leq 40$, $n_s$ values  (black points) computed from Eq.~\eqref{eq: ns analytic} by fixing reheating to its maximum allowed duration, $\Delta N_\text{rh}=\Delta N_\text{rh, max 2}$ in this case.  
    We extrapolate these results to larger values of $p$ by substituting in Eq.~\eqref{eq: ns analytic} $\Delta N_\text{CMB}$ as given in Eq.~\eqref{eq: DN CMB} and the fitting functions~\eqref{eq: rh fit as function of p} and~\eqref{eq: DN CMB inst rh fit as function of p} (dashed, pink line), which we represent together with the $95\%$ C.L. bands for the fit.  
    The black star marks the asymptotic value of $n_s$ for $p\to \infty$.}
    \label{fig: ns for p large}
\end{figure}
The numerical results are displayed in Fig.~\ref{fig: ns for p large} together with the function obtained by substituting the fitting functions~\eqref{eq: rh fit as function of p} and~\eqref{eq: DN CMB inst rh fit as function of p} in Eq.~\eqref{eq: ns analytic}. 
Thanks to the latter, we are able to extrapolate $n_s$ to even larger $p$. 
We obtain the asymptotic result $\lim_{p\to \infty} n_s = 0.968165\approx 0.9682$. 
We have approximated to the fourth decimal figure since the theoretical error in computing $n_s$ is $\lesssim 0.0003$, see Appendix~\ref{app: theoretical errors}. 
If future CMB experiments establish $n_s>0.9682$ at least at $95\%$ C.L. then we will be able to rule out T-models at least at $2\sigma$. 

The \textit{Planck} and BICEP/Keck 2018 experimental error in the $\Lambda\text{CDM}+r$ cosmology is $\sigma(n_s)=0.0041$~\cite{Paoletti:2022anb}, while including ACT data yields $\sigma(n_s)=0.0033$~\cite{ACT:2025tim}. 
Future, ground-based CMB-S4 experiments are designed to achieve $\sigma(n_s)=0.0019$~\cite{CMB-S4:2016ple}, and more advanced surveys will slightly improve the $n_s$ measurement, e.g. the proposed space-based PICO is designed to achieve $\sigma(n_s) = 0.0015$~\cite{NASAPICO:2019thw}. 
These considerations indicate that future surveys will measure the third decimal figure with increasing precision. 
Given these experimental uncertainties, what are the prospects for constraining the $p$ parameter through $n_s$? 
We find that an experiment like ACT cannot distinguish between a model with $p=4$ and $p=6$, as the corresponding $n_s$ values differ by less than $\sigma(n_s)$. 
An advanced survey as PICO will not be able to distinguish between $p=6$ and $p=8$.
    \footnote{To obtain these results we have taken into account that $\Delta N_\text{rh, max}=\Delta N_\text{rh, max 1}$ when $p\leq 6$, see the top right panel in Fig.~\ref{fig: BBN maximum duration of reheating}.}

Let us close with a few comments on the status of T-models with large $p$. 
In this section we explored up to $p=40$, and one might question to which extent these are theoretically motivated models.
To explain the issue, let us consider the formulation of T-models within supergravity. 
They are derived in terms of a complex field $Z\equiv \tanh{\left(\phi/\sqrt{6\alpha}\right)} \,e^{i\theta}$ by assuming a potential which is invariant under a phase-shift $Z\to e^{i\varphi} Z$. 
In terms of the canonical inflaton field $\phi$, the potential reads $V(\phi) = c_2 \tanh^2{\left(\phi/\sqrt{6\alpha}\right)} + c_4 \tanh^4{\left(\phi/\sqrt{6\alpha}\right)} + \cdots$. 
Here we choose to work in the idealised case in which the dominant term in the series is $c_p \tanh^p{\left(\phi/\sqrt{6\alpha}\right)}$, see Eq.~\eqref{eq: T-model potential}.
We assume this to still be true when $|\phi|\ll \sqrt{6\alpha}$, i.e. around the minimum of the potential, and thus we neglect the effect of lower-order powers on the reheating equation of state~\eqref{eq: equation of state parameter as function of p}. 
This is of course an idealised framework when $p>2$, whose robustness should be investigated.
On the other hand, it is \emph{not} necessary to work with very large $p$ values in order to reconcile T-models with the P-ACT-LB-BK18 $n_s$ measurement.
Indeed, we showed in Sec.~\ref{sec: predictions vs PACT} that for $p\geq 6$ the $n_s$ and $r$ predictions enter the $95\%$ C.L. contours already in the most constraining reheating scenario we work with $(T_\text{rh}>1\,\text{TeV})$. 
While exploring the large $p$ limit demonstrates the asymptotic behavior of $n_s$, from an observational point of view T-models are highly predictive in this regime; for example, only the fifth decimal figure of $n_s$ is affected by $p$ in Fig.~\ref{fig: ns for p large}.
Even futuristic CMB measurement will not be able to distinguish between models with $p\geq 6$. 

\section{Discussion}
\label{sec: discussion}

Recent ACT DR6~\cite{ACT:2025fju, ACT:2025tim} and SPT-3G data~\cite{SPT-3G:2025bzu} combined with BAO DESI data show a preference for $n_s$ larger than that determined by \textit{Planck}~\cite{Planck:2018jri}. 
Pending a resolution of the tension between BAO parameters as determined from CMB and
DESI data~\cite{SPT-3G:2025bzu, Ferreira:2025lrd}, these results motivate us to explore the large-$n_s$ regime of inflationary models. 
We choose to focus on $\alpha$-attractor T-models~\eqref{eq: T-model potential}, in which case, values of $n_s$ closer to unity can be produced when $p$ is large.
Indeed, for $p\geq 6$, reheating has a stiff equation of state, see Eq.~\eqref{eq: equation of state parameter as function of p}. 
For extended reheating, $\bar w>1/3$ leads to larger $\Delta N_\text{CMB}$ values thereby pushing $n_s$ closer to unity (see the $n_s$ universal prediction, Eq.~\eqref{eq: ns universal}, for a schematic illustration of this effect).
In Secs.~\ref{sec: inflation and reheating} and~\ref{sec: inflationary predictions} we work in a region of the parameter space defined by even $p\in[2,\,20]$ and $10^{-4}\leq \alpha\leq 20$, and extend further the $p$ range in Sec.~\ref{sec: how far can we push ns in these models}. 

{The monomial T-model potential used in our work, Eq.~\eqref{eq: T-model potential}, provides a minimal setup in which $p$ directly controls both inflationary observables and the effective reheating equation of state. 
We leave the study of more general polynomial functions of $\tanh$ for future work.}

\subsubsection*{Implications of a large-$\bm{n_s}$ measurement for T-models}
In Sec.~\ref{sec: inflationary predictions} we use P-ACT-LB-BK18 data~\cite{ACT:2025tim} to illustrate what a large-$n_s$ measurement would imply for T-models.
As expected, we find that models with matter-dominated $(p=2)$ or instantaneous reheating $(p=4)$ are not compatible with P-ACT-LB-BK18.  
Instead, $n_s$ and $r$ predictions for models with $p\geq 6$ can enter the $95\%$ C.L. P-ACT-LB-BK18 posterior, for a range of theory parameters $\{\alpha,\,\Delta N_\text{rh}\}$ which depends on $p$ and on the limits imposed on the duration of reheating.
In particular, for fixed $\Delta N_\text{rh}$ the range of $\alpha$ is more extended for larger values of $p$. 
Predictions in the $(n_s,\,r)$ plane obtained by requiring that reheating is complete before the universe reaches $1\,\text{TeV}$ $(T_\text{rh}>1\,\text{TeV})$ are displayed in Fig.~\ref{fig:predictions in ns and r plane}, and the resulting weakest constraints in the $(\alpha,\,p)$ parameter space are shown in the top panel of Fig.~\ref{fig: ACT constraints on parameter space}. 
If reheating is allowed to last up to when BBN takes place, for fixed $p$, more $\{\alpha,\Delta N_\text{rh}\}$ models are compatible with P-ACT-LB-BK18 with respect to the  $T_\text{rh}>1\,\text{TeV}$ case. 
To see this, compare the $p=8$ panels in Figs.~\ref{fig:predictions in ns and r plane} and~\ref{fig: BBN predictions in ns and r plane}, or the upper and lower panels of Fig.~\ref{fig: ACT constraints on parameter space} which are obtained for $\Delta N_\text{rh}=\Delta N_\text{rh, max}$.
Regardless of the maximum allowed duration of reheating, we find that $n_s$ is maximised for $\alpha\approx1$, see Fig.~\ref{fig: ACT constraints on parameter space}. 
This shows that the seven Poincar\'e models~\cite{Ferrara:2016fwe, Kallosh:2017ced} are very well placed to deliver large $n_s$. Crucially, it also implies that the tensor-to-scalar ratio cannot lie too far below current observational bounds on $r$.

\subsubsection*{The many effects of $\bm{p}$}
Above we have focused on the effect that $p\geq 6$ $(\bar w>1/3)$ has in pushing $n_s$ closer to unity by increasing the value of $\Delta N_\text{CMB}$, see the third term on the right-hand-side of Eq.~\eqref{eq: DN CMB}.
Nevertheless, increasing $p$ has two more effects on $n_s$, which go in the opposite direction. 

First, for $\alpha\gtrsim \mathcal{O}(1)$ the $p$-dependent corrections in Eq.~\eqref{eq: ns analytic} become important, and they make $n_s$ smaller. 
One can see this effect in Figs.~\ref{fig:predictions in ns and r plane} and~\ref{fig: BBN predictions in ns and r plane}, where for fixed $\{p,\,\Delta N_\text{rh}\}$ and increasing $\alpha$ the $n_s$ and $r$ predictions turn around when $\alpha\sim\mathcal{O}(1)$.
It is thanks to this effect that we are able to place upper limits on $\alpha$, see Fig.~\ref{fig: ACT constraints on parameter space} for the case $\Delta N_\text{rh}=\Delta N_\text{rh, max}$. 
While it is well established that $\alpha$ can be bounded from above by constraining $r$, our results show that including $n_s$ results into a stronger bound.
Note that this effect would be erroneously missed if one were to use the universal predictions~\eqref{eq: ns universal} and~\eqref{eq: r universal}. 

Second, the duration of reheating needed to reach a fixed energy density $\rho_\text{rh}$ is smaller the larger $p$ is, see Eqs.~\eqref{eq: equation of state parameter as function of p} and~\eqref{eq: Delta N rh def}.  
This in turn implies that the maximum allowed duration of reheating is smaller for large $p$, see the left panel of Fig.~\ref{fig: maximum duration of reheating} for $\Delta N_\text{rh, max 1}$ computed assuming $T_\text{rh}>1\,\text{TeV}$.
It follows that, while the coefficient of the third term on the right-hand-side of Eq.~\eqref{eq: DN CMB} is (positive and) larger the larger $p$ is, the duration of reheating $\Delta N_\text{rh}$ is more constrained, thereby limiting the ability of stiff reheating to increase $n_s$.
To assess the interplay between these two effects one can compute $\Delta N_\text{CMB}$~\eqref{eq: DN CMB} for fixed $\alpha$ and maximum duration of reheating, and check whether it always increases for larger $p$. 
In the range $p\in[2,\,20]$ and $10^{-4}\leq \alpha\leq 20$, and for both reheating scenarios ($T_\text{rh}>1\,\text{TeV}$ and $T_\text{rh}>T_\text{BBN}$), we find that for increasing $p$, $\Delta N_\text{CMB}$ either always increases or flattens (the case for models affected by the GW bound~\eqref{eq: Omega GW Neff bound}). 
In other words, the increase in $\bar w$ obtained by increasing $p$ makes $\Delta N_\text{CMB}$ larger, winning over the decrease in $\Delta N_\text{rh, max}$. 

\subsubsection*{Saturation of large-$\bm{p}$ effects}
In the large-$p$ limit, $\bar w$ asymptotes to $1$, see Eq.~\eqref{eq: equation of state parameter as function of p}. 
Due to this, the capability of an extended and stiff reheating phase to increase the compatibility of T-models with P-ACT-LB-BK18 saturates. 
This is shown, e.g., in Fig.~\ref{fig: BBN predictions in ns and r plane} where one sees that there is no appreciable increase in the P-ACT-LB-BK18 $95\%$ C.L. region covered by compatible models when moving from $p=16$ to $p=20$. 
Thanks to this effect, in Sec.~\ref{sec: how far can we push ns in these models} we are able to establish the largest $n_s$ that T-models can produce:
\begin{equation}
    \lim_{p\to\infty} n_s \approx 0.9682 \;. \label{eq:nslimmain}
\end{equation}
Our result shows that T-models are highly predictive in the large-$p$ regime we explored, and provides a theoretical benchmark to be compared with future CMB data. 

\subsubsection*{What could go wrong: theoretical modelling keeps up with observations}
With observations becoming more and more precise, a meaningful comparison against inflationary models 
requires careful theory modelling.

First, reheating must be taken into account, and its contribution separated from that of the inflaton potential, see Eq.~\eqref{eq: DN CMB}.
While an extended and stiff reheating stage increases $n_s$, its duration is subject to precise constraints.
In this work, we require reheating to be complete before the universe reaches a specific temperature, and consider two scenarios with $T_\text{rh}>1\,\text{TeV}$ and $T_\text{rh}>T_\text{BBN}$. 
By requiring reheating to satisfy the GW constraint~\eqref{eq: Omega GW Neff bound}, we also ensure that the resulting blue-tilted primordial GWs do not disrupt BBN. 
This is particularly important when $T_\text{rh}>T_\text{BBN}$, as shown in Fig.~\ref{fig: BBN predictions in ns and r plane}. 
The larger $p$ is, the bigger is the error in estimating the ability of T-models to cover the P-ACT-LB-BK18 posterior if the GW constraint~\eqref{eq: Omega GW Neff bound} is not taken into account, especially for models in the large-$n_s$ regime. 

Second, we have carefully computed the $n_s$ and $r$ inflationary predictions by ensuring that the theoretical error is always at least one order of magnitude below the experimental uncertainty (i.e. $\sigma(n_s)$ in Eq.~\eqref{eq:ns P-ACT-LB-BK18}) for the P-ACT-LB-BK18 combination.
In App.~\ref{app: theoretical error analytical prediction} we focus on the error made by approximating the second-order slow-roll results for $n_s$ with Eq.~\eqref{eq: ns analytic}.
Importantly, we show that truncating $n_s$ at first-order in the slow-roll expansion does not provide results that are precise enough for comparison with P-ACT-LB-BK18 data. 
In App.~\ref{app: theoretical error from w time dependence} we assess the error made by approximating the reheating equation-of-state parameter with the constant in~\eqref{eq: equation of state parameter as function of p}.

\subsubsection*{Comparison with the literature}
Here, we compare our work with others that also considered the impact of reheating  in increasing the compatibility of $\alpha$-attractors with recent CMB observations. 
See Ref.~\cite{Kallosh:2025ijd} and references therein for a comprehensive discussion of how $\alpha$-attractor models can be reconciled with recent measurements. 

In Ref.~\cite{Drees:2025ngb} the authors consider the Starobinsky model~\cite{Starobinsky:1980te}, which belongs to the class of $\alpha$-attractor E-models~\cite{Kallosh:2015zsa} upon setting $\alpha=1$. 
A modified version of the universal predictions~\eqref{eq: ns universal} and~\eqref{eq: r universal} is used to compute $n_s$ and $r$ at the CMB scale.
The authors consider the effect that a stiff $\bar w$ has in reconciling the model with P-ACT-LB-BK18, and their findings are qualitatively in agreement with ours. 
However, the Starobinsky potential is quadratic at the minimum, leading to $\bar w=0$, and additional features are required to support large $\bar w$.
Instead, for the T-models~\eqref{eq: T-model potential} we consider, the potential parameter $p$ controls $\bar w$, see Eq.~\eqref{eq: equation of state parameter as function of p}, and stiff $\bar w$ is produced whenever $p\geq 6$. 

In Ref.~\cite{Haque:2025uri} both E- and T-models are considered, and the equation-of-state parameter is connected to $p$ as in Eq.~\eqref{eq: equation of state parameter as function of p}.
Reheating is constrained to be complete before the onset of BBN, and to satisfy the GW constraint~\eqref{eq: Omega GW Neff bound}. 
The main difference between our work and Ref.~\cite{Haque:2025uri} lies in the computation of the $n_s$ and $r$ predictions. 
In Ref.~\cite{Haque:2025uri} the first-order slow-roll expressions are used (in their formulation, in terms of potential slow-roll parameters). 
As shown in App.~\ref{app: theoretical error analytical prediction}, these are not necessarily ideally suited for comparison with P-ACT-LB-BK18 data. 
See Ref.~\cite{Haque:2025uga} for the case of polynomial $\alpha$-attractors.

In Ref.~\cite{German:2025ide} a generalisation of $\alpha$-attractor models\footnote{See also Ref.~\cite{Alestas:2024eic} for an $\alpha$-attractor-inspired quintessence model which could potentially explain the time-dependent dark-energy equation of state measured by DESI.}
is compared with ACT DR6 data. 
The behavior of the inflationary potential close to the minimum is different from that of T-models, and therefore the equation-of-state parameter is different from Eq.~\eqref{eq: equation of state parameter as function of p}.
In particular, it always features a quadratic term, leading to $\bar w=0$.
Ref.~\cite{German:2025ide} considers $\bar w\neq 0$ and finds that $1/3<\bar w<1$ is required to reach the P-ACT-LB-BK18 $(n_s,\,r)$ posterior.

While the present work was being written up, Ref.~\cite{Ellis:2025zrf} considered T-models and the impact of stiff $\bar w$. 
Whenever the parameter spaces being investigated overlap, our results are qualitatively in agreement with those of Ref.~\cite{Ellis:2025zrf}. Our analysis here goes further in that we implement the GW constraint~\eqref{eq: Omega GW Neff bound} on the duration of reheating. 
We find that, when reheating is allowed to last up to BBN, models with $p\geq 8$ are impacted (see Fig.~\ref{fig: BBN maximum duration of reheating}). It would be interesting to implement the GW constraints in the context presented in Ref.~\cite{Ellis:2025zrf}.
We note that in Ref.~\cite{Ellis:2025zrf} the effect of inflaton fragmentation is taken into account. 
Indeed, when $p>2$ inflaton self-interactions during reheating enhance spatial inhomogeneities, up to a point in which the field is no more coherently oscillating around the potential minimum but rather it is fragmented and the equation of state parameter rapidly approaches $1/3$~\cite{Lozanov:2016hid, Lozanov:2017hjm} (see e.g. Refs.~\cite{Apers:2024ffe, Eroncel:2025bcb, Mosny:2025cyd} for the case of kination, $w=1$).
We plan to incorporate fragmentation bounds on the duration of reheating in our T-models analysis in future work. 

Other than a stiff equation of state during reheating, other effects have been recently considered that could ease the tension between T-models and the new P-ACT-LB-BK18 $n_s$ measurement. 
These include corrections to the inflationary potential due to self-interactions or couplings to matter fields (either included perturbatively~\cite{Wolf:2025ecy, Ellis:2025bzi}, or in a non-perturbative framework~\cite{Alexandre:2025ixz}), non-Bunch--Davies initial conditions~\cite{Maity:2025czp}, a  generalized uncertainty principle~\cite{Heidarian:2025drk}, and Gauss-Bonnet couplings~\cite{Yogesh:2025wak, Zhu:2025twm}. 

{In a broader context, recent work~\cite{Ghoshal:2025ejg} has been exploring the combined use of CMB observables and reheating temperature to arrive at a consistency check between inflationary theories and dark matter production.}

Let us close by highlighting the key results of our work. 
The main novelty of our work lies in the fact that we
(i) investigate the asymptotic behavior in the large-$p$ regime and provide the largest $n_s$ value that T-models can produce;
(ii) explore the multiple effects of the parameter $p$, and point out that the $p$-induced turn-around of the $(n_s,\,r)$ predictions when $\alpha=\mathcal{O}(1)$ allows us to place an upper limit on $\alpha$ that is stronger than the one from $r$ alone;
and
(iii) find that, regardless of the limits on reheating and the value of $p$, $n_s$ is maximised for $\alpha\approx 1$.

\section*{Acknowledgments}
LI is grateful to H. Jense and C. Hill for discussing the P-ACT-LB-BK18 $n_s$ measurement, and to M. Ballardini for a clarification on the difference between second- and third-order slow-roll predictions for $n_s$. 
The authors are grateful to E. Ferreira, L. Heurtier, R. Kallosh and A. Linde for helpful discussions.
This work was supported by the Science and Technology Facilities Council (grant numbers ST/W001225/1 and ST/X000931/1).
SB thanks Cristobal Zenteno for pointing out several interesting works on the topic. SB and MF acknowledge the ``Consolidaci\'{o}n Investigadora'' grant CNS2022-135590. The work of SB and MF is partially supported by the Spanish Research Agency (Agencia Estatal de Investigaci\'{o}n)
through the Grant IFT Centro de Excelencia Severo Ochoa No CEX2020-001007-S, funded by
MCIN/AEI/10.13039/501100011033. MF acknowledges support from the ``Ram\'{o}n y Cajal'' grant
RYC2021-033786-I. Initial discussion for this work was carried out during the 2025 ``The Dawn of Gravitational Wave Cosmology'' workshop, supported by the Fundaci\'{o}n Ram\'{o}n Areces and hosted by the ``Centro de Ciencias de Benasque Pedro Pascual''. We thank both the CCBPP and the Fundaci\'{o}n Areces for creating a stimulating and very productive environment for research.
For the purpose of open access, the authors have applied a Creative Commons Attribution (CC-BY) licence to any Author Accepted Manuscript version arising from this work.
Supporting research data are available on reasonable request from the corresponding author, Laura Iacconi.

\appendix

\section[\texorpdfstring{Theoretical errors in computing $\bm{n_s}$}{Theoretical errors in computing ns}]{Theoretical errors in computing $\bm{n_s}$}
\label{app: theoretical errors}
In this Appendix we discuss the sources of error in the computation of the scalar spectral tilt, $n_s$, and compare them with the current observational uncertainty $\sigma(n_s)$ obtained from the P-ACT-LB-BK18 data combination, see Eq.~\eqref{eq:ns P-ACT-LB-BK18}. 
In App.~\ref{app: theoretical error analytical prediction} we consider the error one makes in computing $n_s$ from Eq.~\eqref{eq: ns analytic} instead of using the corresponding expression at second-order in the slow-roll expansion. 
In App.~\ref{app: theoretical error from w time dependence} we consider instead the time-dependence of the reheating equation-of-state parameter, and evaluate the error made in computing $n_s$ by assuming a constant $w=\bar w$, see Eq.~\eqref{eq: equation of state parameter as function of p}. 

\subsection{Use of analytical expression}
\label{app: theoretical error analytical prediction}
For four illustrative values of $p$, $p=\{2,\, 4,\, 6,\, 8\}$, and $10^{-4}\leq \alpha\leq 20$ we compute $n_s$ by using the analytical expression in Eq.~\eqref{eq: ns analytic} and compare it with the value obtained numerically by using the second-order slow-roll expression.
    \footnote{In Refs.~\cite{Auclair:2022yxs, Ballardini:2024irx} expressions have been derived up to third-order in the slow-roll expansion. 
    It is sufficient here to work with the second-order results because the difference between the two is much smaller than $\sigma(n_s)$ in Eq.~\eqref{eq:ns P-ACT-LB-BK18}~\cite{Ballardini:2024irx}.
    Even considering a futuristic CMB experiment, for a T-model with $\alpha=1$ the first, second and third Hubble-slow-roll parameters are well recovered without any bias or significant change in the uncertainties by stopping at second order~\cite{Ballardini:2024irx}.}
For the latter we employ the expression in terms of potential slow-roll parameters, see Eq.~(B.3) in Ref.~\cite{Iacconi:2023mnw}. 
\begin{figure}
\centering
\captionsetup[subfigure]{justification=centering}
    \begin{subfigure}{.48\textwidth}
        \includegraphics[width=\textwidth]{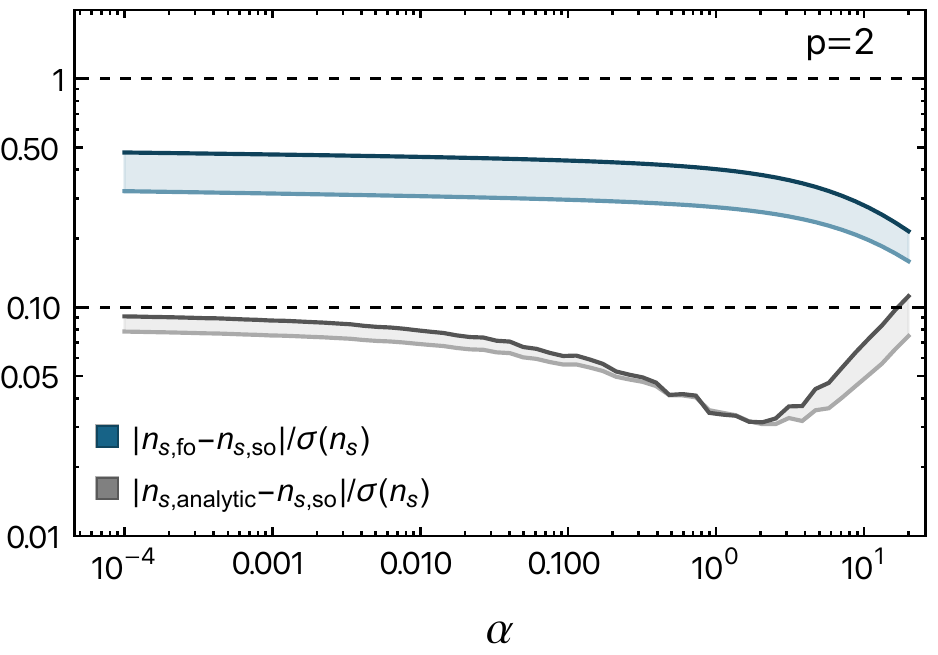}
    \end{subfigure}
    \begin{subfigure}{.48\textwidth}
        \includegraphics[width=\textwidth]{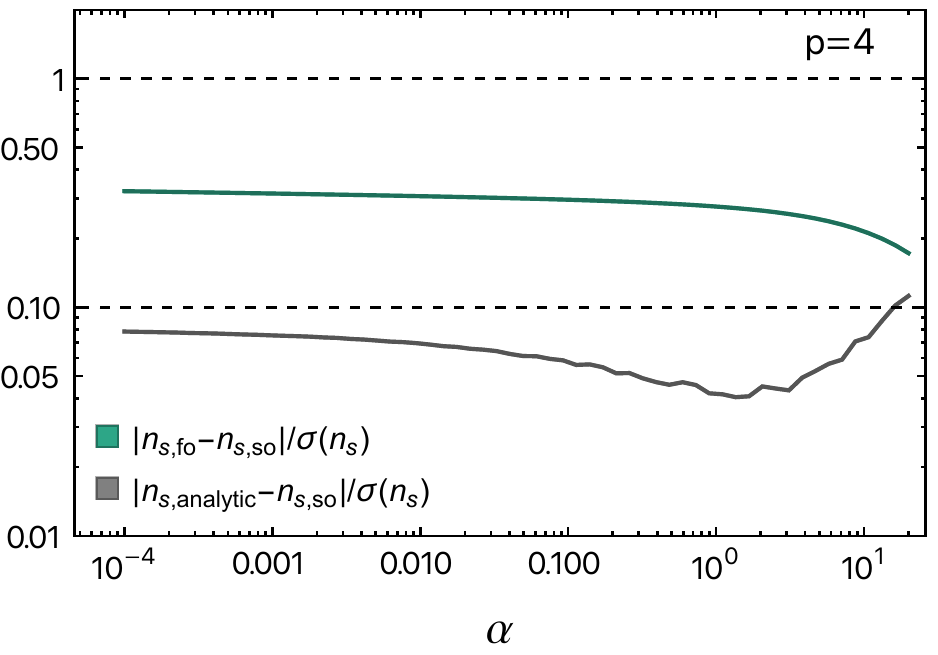}
    \end{subfigure}
    \begin{subfigure}{.48\textwidth}
        \includegraphics[width=\textwidth]{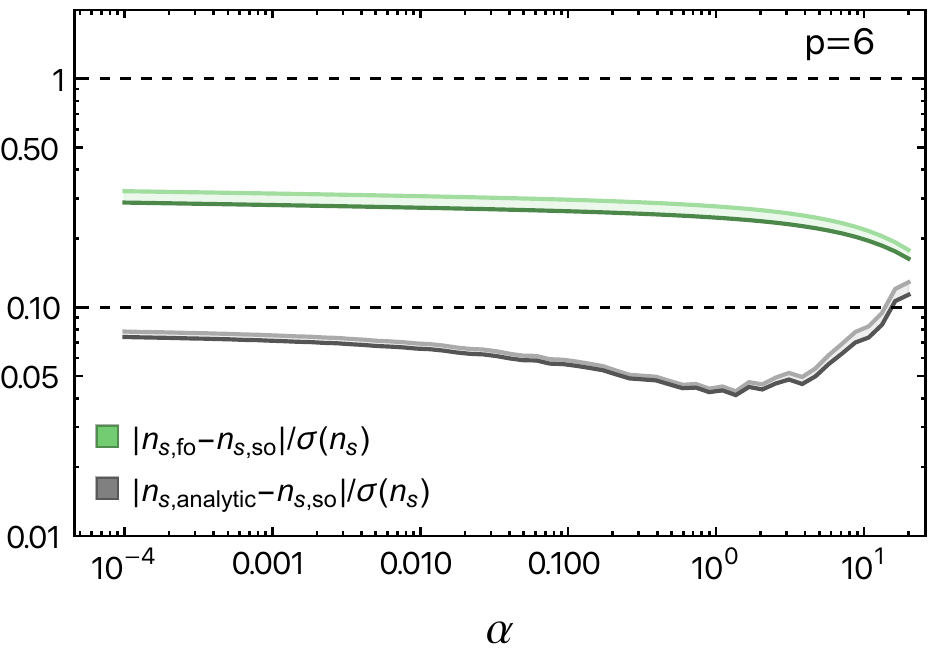}
    \end{subfigure}
    \begin{subfigure}{.48\textwidth}
        \includegraphics[width=\textwidth]{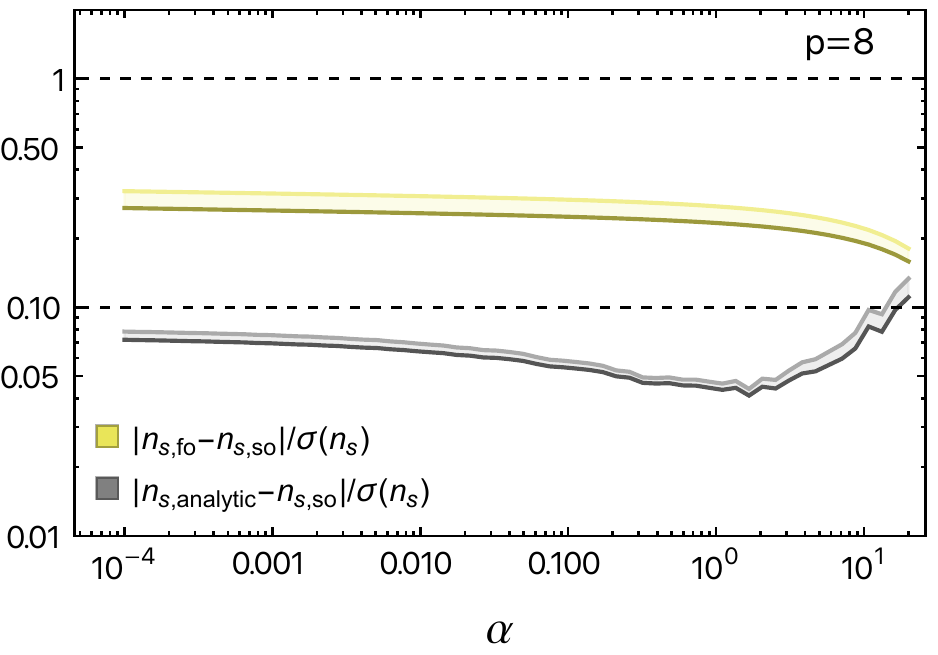}
    \end{subfigure}
\caption{
Difference between $n_s$ computed at first-order in slow-roll (colored band) and with the analytical expression~\eqref{eq: ns analytic} (gray band) with respect to a computation performed at second-order in slow-roll. 
All values are normalised with respect to $\sigma(n_s)$ obtained from P-ACT-LB-BK18~\cite{ACT:2025tim}, see Eq.~\eqref{eq:ns P-ACT-LB-BK18}.  
The two dashed, horizontal lines indicate when the difference in $n_s$ is $100\%$ and $10\%$ of $\sigma(n_s)$ respectively. 
The extended bands include all possible reheating scenarios produced by requiring $T_\text{rh}>1\,\text{TeV}$ (see Sec.~\ref{sec: duration of reheating}), with $\Delta N_\text{rh}=0$ ($\Delta N_\text{rh}=\Delta N_\text{rh, max 1}$) corresponding to the light (dark) limiting line.
Each panel corresponds to a different value of $p$. 
}
     \label{fig:theoretical ns error first vs second order slow roll}
\end{figure}
The results are shown in gray in Fig.~\ref{fig:theoretical ns error first vs second order slow roll} for the whole range of allowed $\Delta N_\text{rh}$ such that $T_\text{rh}>1\,\text{TeV}$, see Sec.~\ref{sec: duration of reheating}. 
Whilst the exact difference between Eq.~\eqref{eq: ns analytic} and the second-order slow-roll $n_s$ is model dependent, for the range of parameters considered this is always at least one order of magnitude smaller than the experimental $\sigma(n_s)$. 
This  ensures that Eq.~\eqref{eq: ns analytic} is sufficiently accurate  to be consistently used when comparing T-models with P-ACT-LB-BK18 data. 

In Fig.~\ref{fig:theoretical ns error first vs second order slow roll} we also show the difference between $n_s$ computed at first-order and second-order in slow-roll, see Eq.~(B.3) in Ref.~\cite{Iacconi:2023mnw}. 
In this case, the difference is comparable to $\sigma(n_s)$, being broadly of the same order of magnitude. 
These results indicate that one should \emph{not} compute predictions for $n_s$ by employing the first-order slow-roll expression, as the theoretical error made is comparable with the experimental one.
We also note that at first-order in slow-roll the difference between the values computed using the expression in terms of Hubble slow-roll parameters or in terms of potential slow-roll parameters is $\mathcal{O}(10^{-3})$, introducing arbitrariness in the choice of which expression to use~\cite{Iacconi:2023mnw}. 
On the other hand, at second-order the difference is one order of magnitude smaller~\cite{Iacconi:2023mnw}. 

The different performance of the analytical prediction in Eq.~\eqref{eq: ns analytic} with respect to the first-order potential slow-roll expression, $n_s = -6 \epsilon_V + 2\eta_V$, might be puzzling, as the former is in fact derived from the latter. 
Nevertheless, while the first-order slow-roll results in Fig.~\ref{fig:theoretical ns error first vs second order slow roll} are evaluated via substitution of the value of $\phi(\Delta N_\text{CMB})$ extracted from the numerical solution, Eq.~\eqref{eq: ns analytic} is obtained by using the slow-roll analytical solutions for $\phi(\Delta N_\text{CMB})$ and $\phi_\text{end}$.  
We have checked that the slow-roll and numerical results for $\phi(\Delta N_\text{CMB})$ can differ up to $1\%$, with the specific value being model dependent, and that this discrepancy explains the difference between Eq.~\eqref{eq: ns analytic} and the semi-numerical first-order result. 
The fact that the former is closer to the second-order result than the latter is a serendipitous accident.

\subsection[\texorpdfstring{Impact of ${\bar w}$ time-dependence}{Impact of w time-dependence}]{Impact of $\bm{\bar w}$ time-dependence}
\label{app: theoretical error from w time dependence}

For $\alpha=1$ and two illustrative values of $p$, $p=\{2,\, 10\}$, we numerically solve the background equations of motion, Eqs.~\eqref{eq: single field inflaton eom} and~\eqref{eq: phi dot with H dot}, up to 3 e-folds after the end of inflation. 
We then numerically compute the instantaneous equation-of-state parameter, 
\begin{equation}
    \label{eq: w instantaneous}
    w \equiv \frac{P}{\rho} = \frac{1/2 \, \dot \phi^2 -V(\phi)}{1/2 \, \dot \phi^2 +V(\phi)}\;,
\end{equation}
where we label the pressure as $P$ to distinguish it from the potential parameter $p$. 
\begin{figure}
\centering
\captionsetup[subfigure]{justification=centering}
    \begin{subfigure}{.48\textwidth}
        \includegraphics[width=\textwidth]{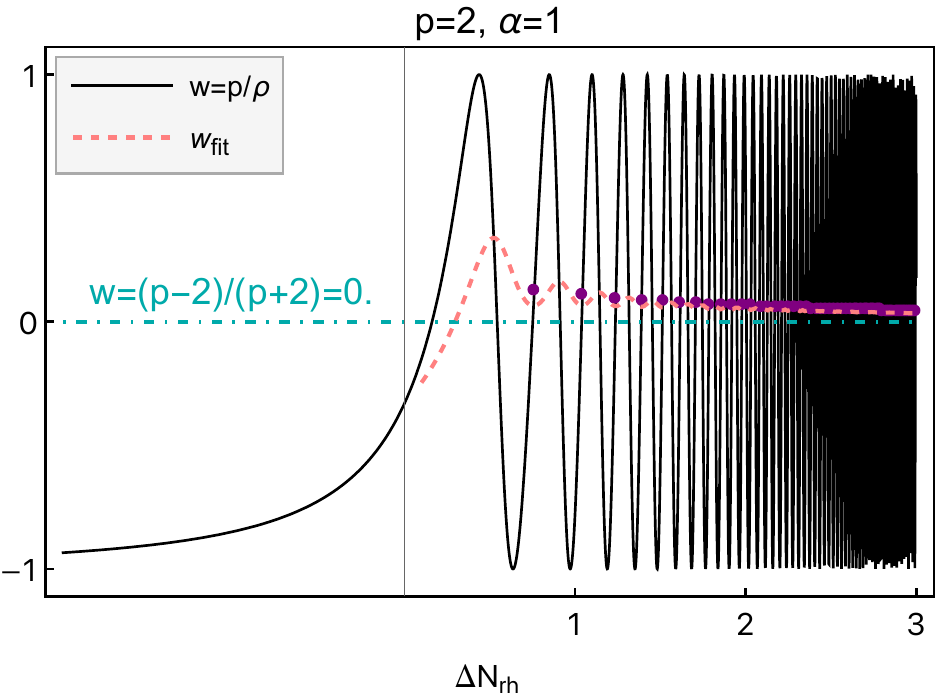}
    \end{subfigure}
    \begin{subfigure}{.48\textwidth}
        \includegraphics[width=\textwidth]{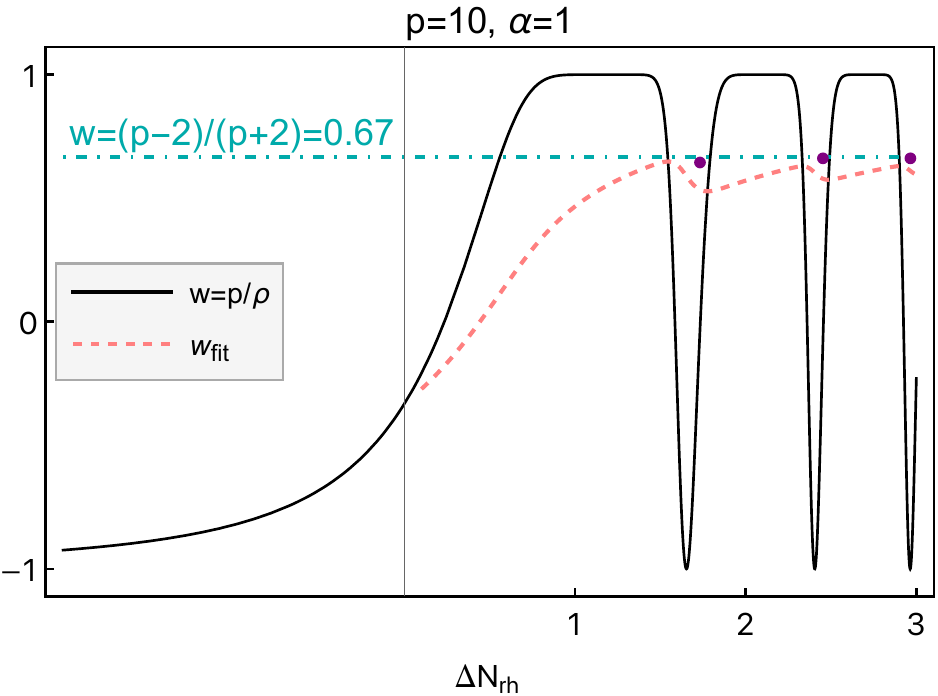}
    \end{subfigure}
\caption{Time-evolution of the equation-of-state parameter, $w$, for two models with $p=2$ (left panel) and $p=10$ (right panel), and $\alpha=1$. 
The thin, vertical line marks the end of inflation, and $\Delta N_\text{rh}$ is the number of e-folds during reheating measured from the end of inflation. 
We represent $w$ as computed by solving numerically the background equations of motion (black line), the averaged quantity $\langle w \rangle$ (purple points) and the fitted values $w_\text{fit}(\Delta N_\text{rh})$ (pink, dashed line). 
The horizontal dot-dashed line marks the value of $\bar w$ computed from Eq.~\eqref{eq: equation of state parameter as function of p}.
}
     \label{fig:numerical w}
\end{figure}
Results are shown in Fig.~\ref{fig:numerical w}.
As expected, $w\approx -1$ during inflation, and then displays an oscillatory behavior that tracks the inflaton oscillations. 
On average $w$ asymptotes to Eq.~\eqref{eq: equation of state parameter as function of p}, where 
\begin{equation}
    \langle w(t) \rangle \equiv T^{-1} \int_{t}^{t+T}\mathrm{d}t'\,  w(t') 
\end{equation}
and $T$ is the period of a single oscillation. 
In particular, we numerically evaluate $\langle w \rangle$ by integrating Eq.~\eqref{eq: w instantaneous} over the period of one (first purple point), two (second purple point), \textit{etc.} full oscillations. 
In other words, we define the periods $T_i$, each one used to produce one of the purple points, as $T_i \equiv N(w=0|_{(1+2i)^\text{th}\,\text{time}})- N(w=0|_{1^\text{st}\,\text{time}}) $. 

Given a specific numerical background evolution and a value of $\Delta N_\text{rh}$, the effective equation-of-state parameter to be used in Eq.~\eqref{eq: DN CMB} should be computed for $\Delta N_\text{rh}$ by fitting Eq.~\eqref{eq: Delta N rh def} with the numerical values of $\rho_\text{end}\equiv \rho(N_\text{end})$ and $\rho_\text{rh}\equiv \rho(N_\text{rh})$ substituted in. 
We perform this fitting procedure for $0<\Delta N_\text{rh}\leq 3$, label the result as $w_\text{fit}(\Delta N_\text{rh})$ and show it together with $w$ and $\langle w \rangle$ in Fig.~\ref{fig:numerical w}.
The quantity $w_\text{fit}(\Delta N_\text{rh})$ oscillates, with smaller and smaller amplitude as  $\Delta N_\text{rh}$ becomes larger. 
While our results show that the longer reheating is the closer $w_\text{fit}(\Delta N_\text{rh})$ is to $\bar w$, when reheating is very brief $(\Delta N_\text{rh}\lesssim 1)$ the two can differ substantially.

The time-dependence of $w_\text{fit}(\Delta N_\text{rh})$ and its appreciable difference from $\bar w$ for very short reheating raise the question of how large the theoretical error made in computing $n_s$ by using the constant $\bar w$ instead of $w_\text{fit}(\Delta N_\text{rh})$ is. 
\begin{figure}
\centering
\captionsetup[subfigure]{justification=centering}
    \begin{subfigure}{.48\textwidth}
        \includegraphics[width=\textwidth]{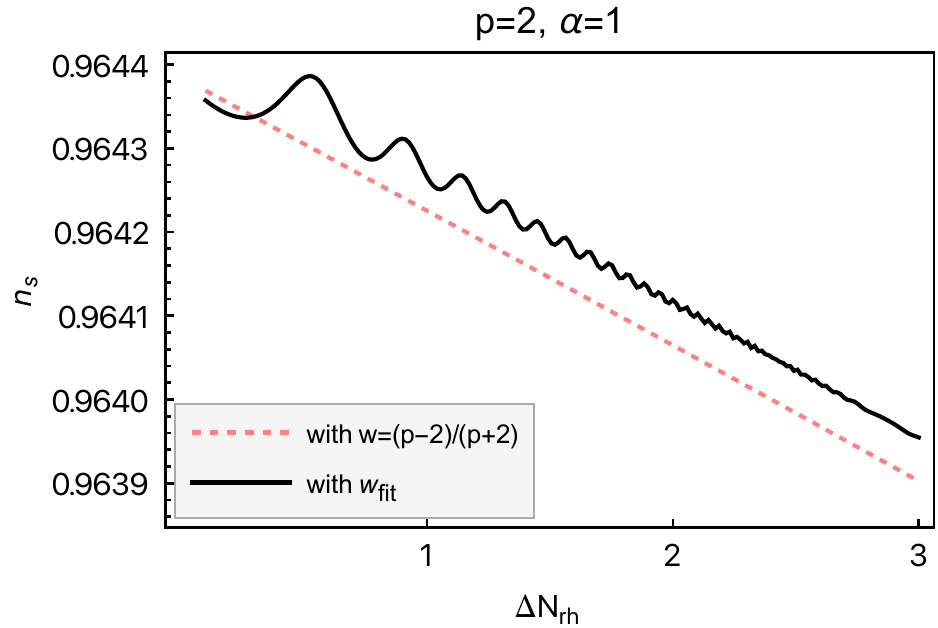}
    \end{subfigure}
    \begin{subfigure}{.48\textwidth}
        \includegraphics[width=\textwidth]{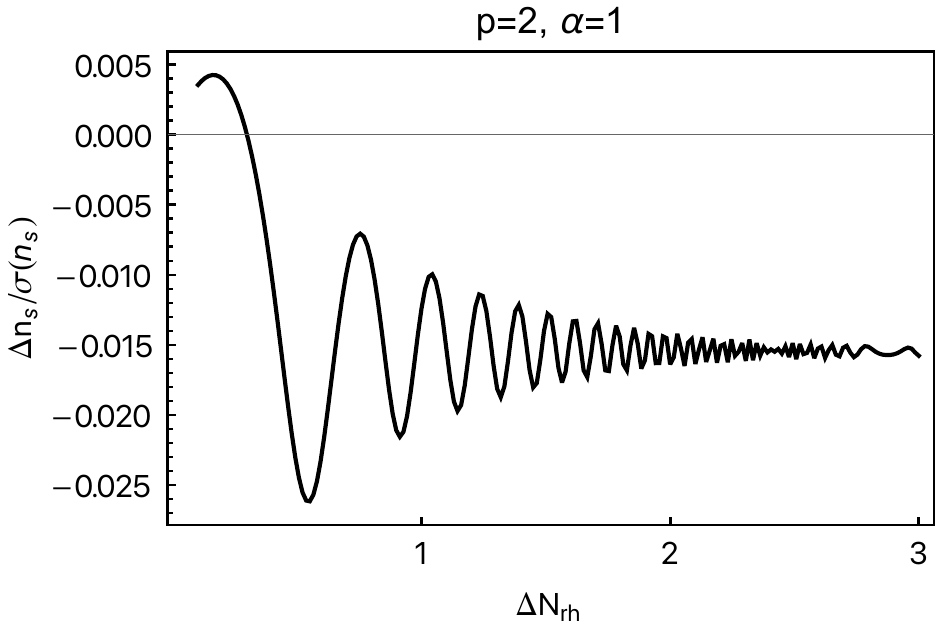}
    \end{subfigure}
    \begin{subfigure}{.48\textwidth}
        \includegraphics[width=\textwidth]{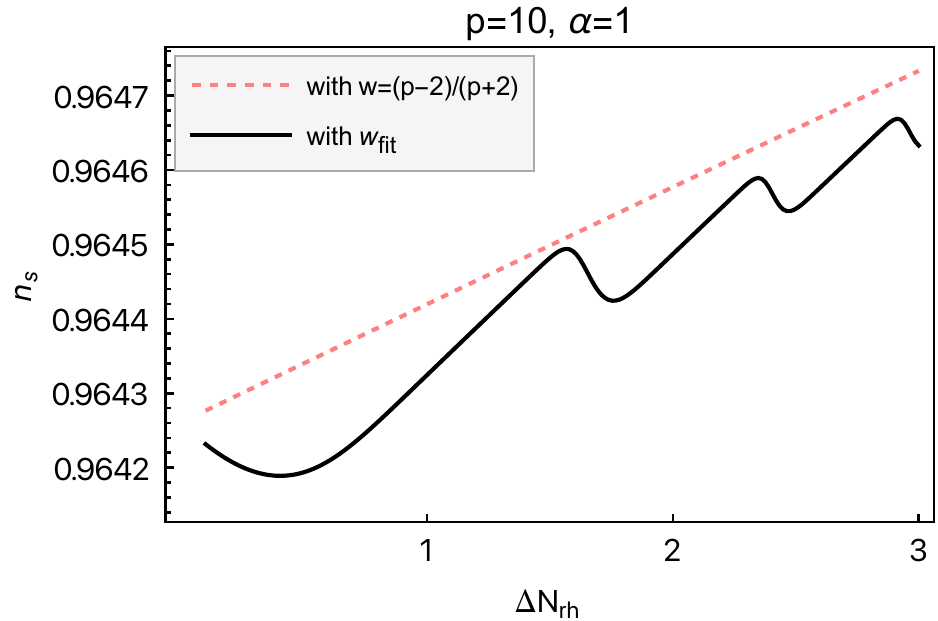}
    \end{subfigure}
    \begin{subfigure}{.48\textwidth}
        \includegraphics[width=\textwidth]{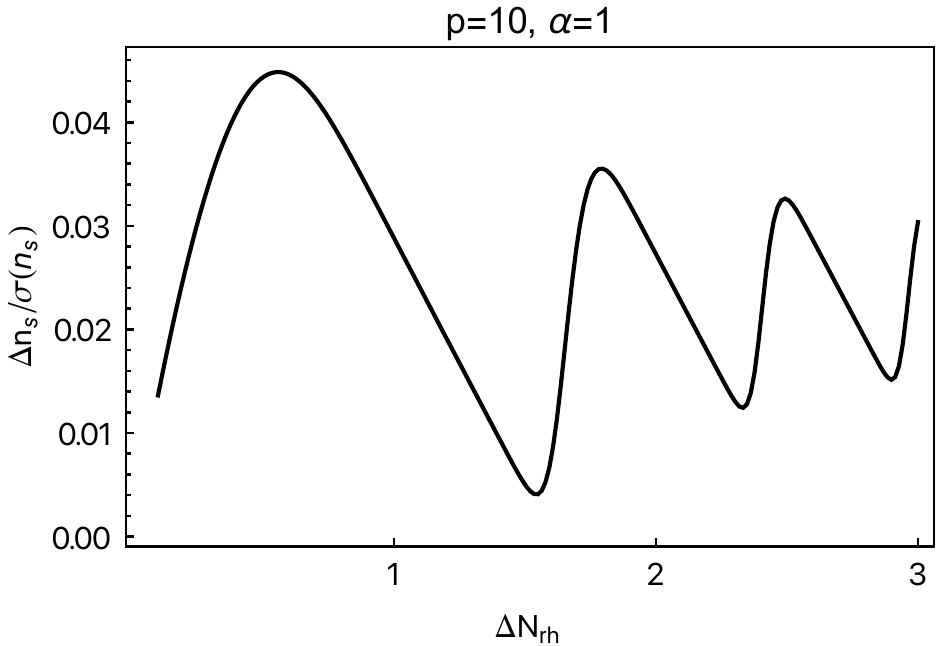}
    \end{subfigure}
\caption{
Scalar spectral tilt, $n_s$, computed from Eq.~\eqref{eq: ns analytic} and represented as a function of the duration of reheating, $\Delta N_\text{rh}$, for the same models displayed in Fig.~\ref{fig:numerical w}. 
\textit{Left panels:} $n_s$ values obtained by fixing the reheating equation-of-state parameter to the value given in Eq.~\eqref{eq: equation of state parameter as function of p} (dashed, pink line) and $n_s$ values obtained by using $w_\text{fit}(\Delta N_\text{rh})$ (continuous, black line). 
\textit{Right panels:} difference between the two, $\Delta n_s \equiv n_s|_{w=\bar w} - n_s|_{w=w_\text{fit}(\Delta N_\text{rh})}$, normalised to $\sigma(n_s)$ obtained from P-ACT-LB-BK18 data, see Eq.~\eqref{eq:ns P-ACT-LB-BK18}.}
     \label{fig:theoretical ns error from w}
\end{figure}
For the two illustrative models displayed in Fig.~\ref{fig:numerical w}, we compute $n_s$ by following the same procedure described in Sec.~\ref{sec: how ns and r are computed}, and compare these results with those obtained by using $w_\text{fit}(\Delta N_\text{rh})$ instead, see Fig.~\ref{fig:theoretical ns error from w}. 
The scalar spectral tilt computed with $w_\text{fit}(\Delta N_\text{rh})$ inherits the oscillations of the equation-of-state parameter, with decreasing amplitude the larger $\Delta N_\text{rh}$ is. 
Once the oscillations have decayed away, $n_s$ settles on a different value with respect to the one computed by using $\bar w$. 
This is due to the fact that during the very early stages of reheating, the time-evolution of $\rho$ is not well described by Eq.~\eqref{eq: Delta N rh def}, and this introduces an offset in $n_s$. 
When comparing the theoretical error made by using $\bar w$ instead of $w_\text{fit}(\Delta N_\text{rh})$ to the experimental one from P-ACT-LB-BK18 data we find that this is always within $\mathcal{O}(1)\%$ of $\sigma(n_s)$. 
This is comparable with the theoretical error discussed in App.~\ref{app: theoretical error analytical prediction}, and it is well below the experimental one. 
For this reason, we neglect the time-evolution of $w$ in the main text when computing inflationary predictions.

\bibliography{refs} 
\bibliographystyle{JHEP}

\end{document}